\documentclass[10pt,twocolumn,twoside]{IEEEtran}
\usepackage{amssymb}
\usepackage{amsmath,amsthm}
\usepackage{graphicx}
\usepackage{dblfloatfix} 
\usepackage{multicol}
\usepackage{lipsum}
\usepackage{algorithm}
\usepackage{algpseudocode}
\usepackage{booktabs}
\usepackage{tabularx}
\usepackage{ragged2e}
\usepackage{csquotes}
\usepackage{helvet}
\usepackage{textcomp}
\usepackage{subcaption}
\usepackage[T1]{fontenc}
\usepackage{bm}
\usepackage{verbatim} 
\usepackage{mathtools}
\usepackage{caption}
\usepackage{graphicx} 
\usepackage{threeparttable}
\usepackage{mathrsfs,float} 
\usepackage{setspace,esvect}
 \usepackage{enumitem}
  \newcolumntype{P}[1]{>{\centering\arraybackslash}p{#1}}
  \newcolumntype{M}[1]{>{\centering\arraybackslash}m{#1}}
\usepackage{color}
\usepackage{bm}
\usepackage{mathrsfs,float} 
\usepackage{epstopdf}
\usepackage{float}
\usepackage{gensymb,stmaryrd}
\usepackage{algorithmicx}
 \usepackage{enumerate}
 \usepackage{enumitem}
  \newcolumntype{P}[1]{>{\centering\arraybackslash}p{#1}}
  \newcolumntype{M}[1]{>{\centering\arraybackslash}m{#1}}

\usepackage{hyperref}
\usepackage{textcomp}
\usepackage[normalem]{ulem}
\usepackage{xr}
\usepackage{cite}
\usepackage{caption}

\usepackage{array}
\usepackage{fixltx2e}
\usepackage{url}
\usepackage{multirow}
\begin{document}
\renewcommand{\thepage}{\arabic{page}}
\title{Unlimited Dynamic Range Analog-to-Digital Conversion }
\author{Adithya Krishna$^*$, Sunil Rudresh$^*$, Vishal Shaw$^*$, Hemanth Reddy Sabbella, Chandra Sekhar Seelamantula,~\IEEEmembership{Senior Member,~IEEE}, and Chetan Singh Thakur,~\IEEEmembership{Senior Member,~IEEE}
\thanks{The authors are with the Indian Institute of Science, Bangalore - 560012, India. A. Krishna,  H. R. Sabbella, and C. S. Thakur are with the Department of Electronic Systems Engineering (Email: \{adithyaik, hemanthrs, csthakur\}@iisc.ac.in, Tel.: +91 9731680833); V. Shaw is with Department of Electrical Communication Engineering (Email: vishalshaw@iisc.ac.in, Tel.: +91 8981252530); S. Rudresh and C. S. Seelamantula are with the Department of Electrical Engineering   (Email: \{sunilr,css\}@iisc.ac.in; Tel.: +91 80 22932695, Fax: +91 80 23600444). 

This work was funded by Pratiksha Trust, Indian Institute of Science.
\newline $^*$ Equal contribution. }}

\maketitle
\begin{abstract}
Analog-to-digital converters (ADCs) provide the link between continuous-time signals and their discrete-time counterparts, and the Shannon-Nyquist sampling theorem provides the mathematical foundation. Real-world signals have a variable amplitude range, whereas ADCs, by design, have a limited input dynamic range, which results in out-of-range signals getting clipped. In this paper, we propose an unlimited dynamic range ADC (UDR-ADC) that is based on the modulo operation ({\it \textbf {self-reset}} feature) to alleviate the problem of clipping. The self-reset feature allows for wrapping of the input amplitudes, which preserves the input dynamic range. We present the signal model and a reconstruction technique to recover the original signal samples from the modulo measurements. We validate the operation of the proposed ADC using circuit simulations in $\mathbf{65}$ nm complementary metal-oxide-semiconductor (CMOS) process technology. The validation is supplemented by a hardware prototype designed using discrete components. A performance assessment in terms of area, power requirement, and the signal-to-quantization-noise ratio (SQNR) shows that the UDR-ADC outperforms the standard ones.
\end{abstract} 

\begin{IEEEkeywords}
Analog-to-digital converter (ADC), modulo sampling, self-reset ADC, high-dynamic-range ADC, folding ADC, quantization.
\end{IEEEkeywords}

\IEEEpeerreviewmaketitle
\section{Introduction}
\begin{figure}[t]
\centering
\includegraphics[width=3.5 in]{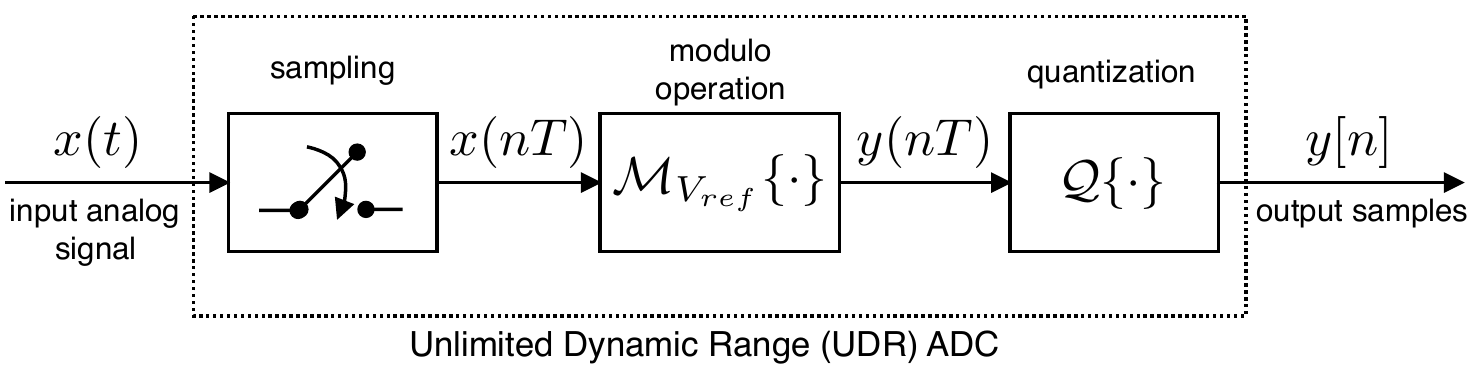}
\caption{Block diagram of a UDR-ADC.}
\label{fig:block_diagram}
\end{figure}

\begin{figure}[t]
\centering
\includegraphics[width=3.5 in]{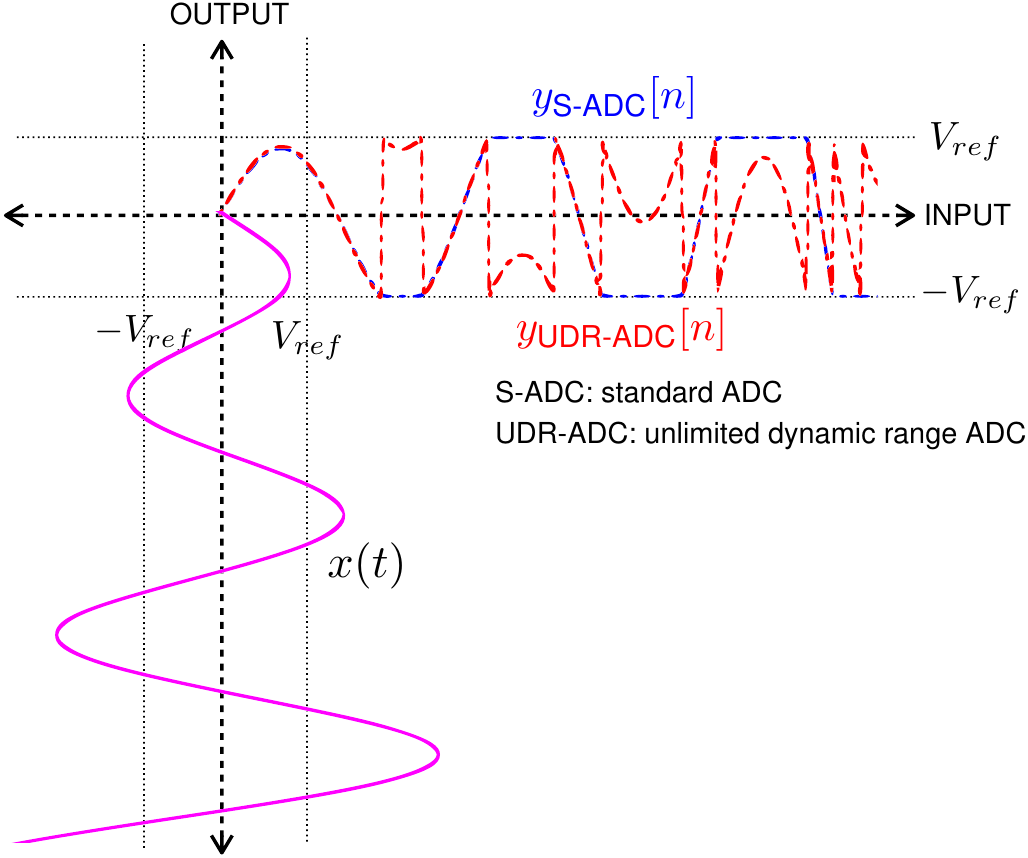}
\caption{[Color online] Input-output characteristics of a UDR-ADC. A standard ADC results in clipping ($y_{\text{S-ADC}}[n]$, shown in blue) whenever the input $x(t)$ goes beyond its dynamic range, whereas a UDR-ADC folds it back ($y_{\text{UDR-ADC}}[n]$, shown in red), thus preserving the input signal up to a modulo.}
\label{fig:ADC_trans}
\end{figure}

Analog-to-digital converters (ADCs) are ubiquitous in signal processing and imaging applications. An ADC embodies the Shannon-Nyquist sampling theorem \cite{shannon}, which provides a fundamental link between continuous-time signals and their discrete counterparts. A crucial parameter that limits the performance of an ADC is the mismatch between the signal dynamic range and the ADC dynamic range (DR). The dynamic range, expressed in dB, is the ratio of the largest signal amplitude to the smallest detectable signal amplitude that the ADC can accurately resolve. A weak signal might get lost in the quantization noise whereas a large signal would drive an ADC into saturation, which clips the signal to a level determined by the power rails. Clipping is a serious problem as it is nonlinear, noninvertible, and severely degrades the spectral content and quality of the digitized signal \cite{Esqueda}. Although performance metrics such as sampling rate, power dissipation, and resolution have been optimized extensively \cite{Walden,Harpe}, an effective solution to the clipping problem has remained relatively unexplored. The objective of this paper is the development of a novel ADC architecture that overcomes this limitation. The proposed ADC takes {\it modulo} measurements whenever the signal goes out of range. In principle, such a design would have an unlimited dynamic range (UDR), from which the name of the proposed ADC is derived. A block diagram is shown in Fig.~\ref{fig:block_diagram}, wherein $V_{ref}$ is the reference voltage against which the modulo is computed, both on the positive and negative voltage swings --- this process is referred to as {\it modulo sampling} in this paper. Figure~\ref{fig:ADC_trans} shows the input-output characteristics of the modulo sampler and also illustrates the advantage of the UDR.\\
The modulo operation on a sampled input analog signal $x(nT)$, $T$ being the sampling interval, is expressed as
\begin{align}
y(nT)&=\mathcal{M}_{V_{ref}}\{x(nT)\}, \text {where}\\
\mathcal{M}_{V_{ref}}\{x(nT)\}&:=\text{mod}\left(x(nT)+V_{ref},2V_{ref}\right)-V_{ref}.
\label{eq:mod_op}
\end{align}
Subtraction of $V_{ref}$ in \eqref{eq:mod_op} ensures that the modulo signal is centered about zero. The signal $y(nT)$ is quantized to obtain the samples $y[n]$: $$y[n]:=\mathcal{Q}\{y(nT)\}=\mathcal{Q}\left\{\mathcal{M}_{V_{ref}}\{x(nT)\}\right\},$$  where $\mathcal{Q}$ denotes the quantization operator. Effectively, the modulo {\it folds} the signal back to the quantizer dynamic range.\\
\indent Before proceeding further, we review the state-of-the-art ADCs that incorporate the folding feature. Such ADCs are referred to as {\it folding} ADCs, self-reset ADCs, or modulo ADCs.

\subsection{State of the Art}
\label{sec:literature}
The early self-reset ADCs were developed primarily for image sensing applications by incorporating pixel-level self-reset either by using a comparator or a Schmitt trigger circuit \cite{ADC_wide_dynamic_range_pixel, SRADC_brain_imaging, CMOS_Imaging, SR_image_sensor, bermak2002vision, zhao2015unbounded}. Rhee and Joo \cite{ADC_wide_dynamic_range_pixel} proposed a design for high-dynamic-range (HDR) imaging with the reset feature and a time-to-digital converter that also increases the peak signal-to-noise ratio (SNR). The CMOS sensor in their design consists of a capacitor for each pixel that integrates the photocurrent. Whenever the voltage of the capacitor reaches a predefined threshold, the capacitor discharges to its initial voltage and starts charging again due to the accumulation of photons. A counter keeps track of the number of resets of a particular pixel. Sasagawa et al.  \cite{SRADC_brain_imaging} demonstrated the application of a CMOS image sensor with the self-reset feature for functional brain imaging. Yuan et al. \cite{CMOS_Imaging} proposed a variant that uses column-wise ADCs instead of pixel-level ones in order to reduce the area and complexity. Since the signal acquisition is based on the photodiode current, such self-reset ADC architectures are specifically applicable for image sensing and the frame rate is limited by the capacitor charge/discharge times. Further, in order to undo the reset operation, one is required to transmit the number of resets at a pixel level. If the number of resets is large, transmitting the side-information would be a significant overhead. In contrast, the proposed architecture keeps track of the reset information using only two bits for encoding three possible states: positive reset, negative reset, or no reset.

General purpose folding ADCs have been introduced in \cite{moreland19958, van19878, van1979high, fiedler1979high} wherein $m$ out of $n$ bits are used to count the number of foldings and they constitute the most significant bits (MSBs). The remaining $n-m$ bits encode the residual voltage and constitute the least significant bits (LSBs). A review of various folding ADCs is available in the article by Kester \cite{folding_ADC}. Folding ADCs require the input signal to be held constant for the entire duration of conversion thus limiting the sampling rate. For a given number of bits, a folding ADC does not provide a dynamic range advantage over a standard ADC. Recently, Ordentlich et al. \cite{ordentlich2018modulo} proposed a phase-based modulo ADC as an alternative to $\Sigma-\Delta$ converters with the goal of reducing the number of quantization levels and thereby the number of bits used for encoding. However, they did not demonstrate any circuit-level feasibility or implementation.

\subsection{Our Contribution}
\label{sec:contribution}
We propose a generic UDR-ADC architecture that incorporates the self-reset feature by means of a custom-designed modulo circuit. The achievable dynamic range in practice is limited not by the quantization block, but only by the voltage supply to the sample-and-hold and modulo circuits, which could be as high as allowed by the process technology. Our contributions are on the design and validation front, and quantization noise analysis.\\
\indent To begin with, we present the signal model and a principled method to reconstruct the signal from the modulo measurements (Section~\ref{sec:reconstruction}). The proposed UDR-ADC consists of three main blocks: (i) a sample-and-hold (S/H) circuit; (ii) a modulo circuit that wraps the input to the predefined range $[-V_{ref}, V_{ref}]$; and (iii) the quantization circuit. Blocks (i) and (iii) are part of any standard ADC. In principle, a standard ADC could be converted into a UDR-ADC by plugging the modulo circuit in between Blocks (i) and (iii). The details of the design are presented in Section~\ref{sec:architectures}. A counter keeps track of the reset information. However, it is not necessary to transmit the counter bits. The counter encodes the reset information in only two bits assuming that the signal does not change by more than $\pm2V_{ref}$ between two consecutive sampling instants. This assumption is reasonable and is directly related to the growth rate of the input signal and the sampling rate. This also significantly reduces the A/D conversion time since the counter can make use of the previous sample reset information. Further, circuit simulation results and a hardware prototype made of discrete components employing the quantization module of a successive approximation register (SAR)-ADC are also presented (Section~\ref{sec:results}) as proof of concept.\\
\indent The advantages of a UDR-ADC in terms of the area and power requirements are analyzed in Section~\ref{sec:performance}. The power consumption of the proposed architecture is significantly less compared with standard ADCs while maintaining nearly the same area. A performance assessment in terms of the signal-to-quantization-noise ratio (SQNR) considering three input signal distributions (uniform, Gaussian, and Laplacian) shows that there is a large operating range where the UDR-ADC outperforms the standard ones  (Section~\ref{sec:SQNR}).

\section{Signal Model for Modulo Sampling}
\label{sec:reconstruction}
 \begin{figure}[t]
\centering
\includegraphics[width=3.5 in]{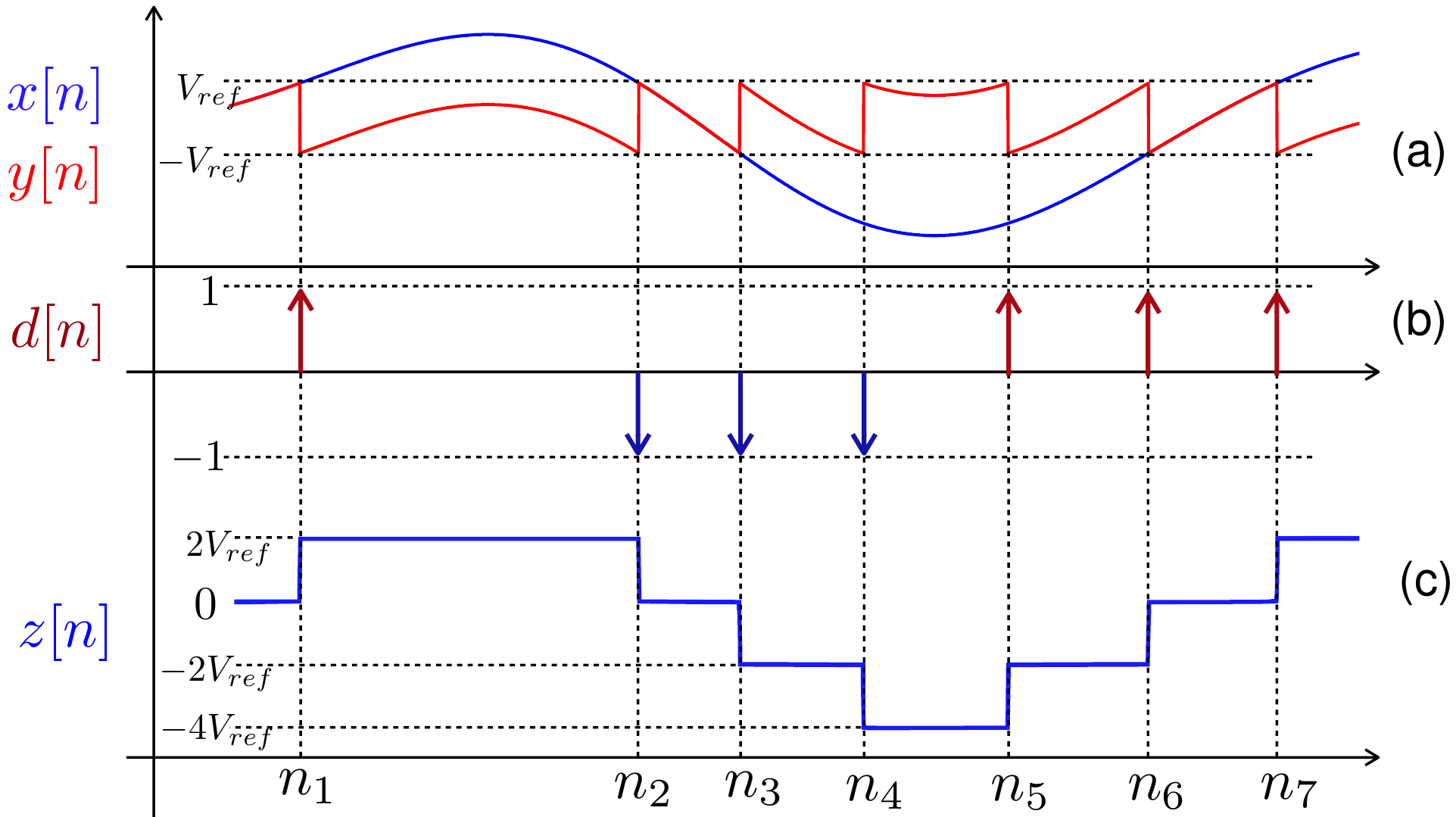}
\caption{[Color online] Illustration of reconstruction of the original signal from the modulo measurements and reset information: (a) original signal samples $x[n]$ and the corresponding modulo samples $y[n]$; (b) reset information $d[n]$; and (c) the discrete-time piecewise-constant signal obtained by a cumulative sum of $d[n]$ and the reset instants $\{n_k\}$. The various sequences are related as $x[n]=y[n]+z[n]$. The sequences are in discrete-time domain, but their plots are shown in a  continuous fashion to aid readability.}
\label{fig:recon_demo}
\end{figure} 
Let $x[n]$ denote the samples of the input signal $x(t)$ and let $y[n]$ denote the modulo measurements. The sequence $x[n]$ can be expressed as the sum of $y[n]$ and a discrete-time piecewise-constant signal $z[n]$:
\begin{align}
x[n] &= y[n] + z[n], \label{eqn:yxz_disc}\\
z[n] &= \sum_k \alpha_k \mathbf{1}_{[n_k,n_{k+1}]}[n],
\label{eqn:z_PCdisc}
\end{align}
where $z[n]$ is an integer multiple of $2V_{ref}$ and $\mathbf{1}_{[n_1,n_{2}]}$ is the discrete-time indicator of the interval $[n_1, n_2]$. The reset instants $\{n_k\}_{k \in \mathbb{Z}}$ are ordered, i.e., $n_k > n_{k-1}$. Figure~\ref{fig:recon_demo} illustrates the signal model, wherein the reset information is encoded in $d[n]$, which is a series of impulses. The piecewise-constant signal $z[n]$ is obtained by accumulating $d[n]$ as $z[n]=2V_{ref}\displaystyle\sum_{k=-\infty}^n d[k]$.\\
\indent Developing reconstruction algorithms for modulo sampling caught the attention of the signal processing community starting from the recent work of Bhandari et al. \cite{unlimited_sampling_ayush, ayush_US_sparse, ayush_US_sinusoid}. We developed a robust reconstruction algorithm using wavelets based on a certain regularity assumption on the ground-truth signal \cite{unlimited_sampling_sunil}. These approaches, despite their robustness and accuracy of reconstruction, require a significantly oversampled input. The reconstruction algorithms are capable of estimating $\{n_k\}$ from $y[n]$ alone provided that $x(t)$ is sufficiently oversampled, i.e., the reset information is not encoded separately. On the contrary, in the proposed UDR-ADC architecture, the oversampling requirement is relaxed to a large extent as the reset information is separately encoded. At every sampling instant, the input is either within $[-V_{ref}, +V_{ref}]$ or outside of it. Whenever it goes outside of $[-V_{ref}, +V_{ref}]$, it could be doing so either with a positive slope or a negative slope. These three possibilities are encoded as: no reset, positive reset, or a negative reset. With respect to Fig.~\ref{fig:recon_demo}(b), the positive resets occur at $n_1,\, n_5,\, n_6$ and $n_7$, whereas the negative ones occur at $n_2,\,n_3,$ and $n_4$. Every sample has two bits dedicated for encoding the reset information and the remaining $n-2$ bits are used for quantizing the amplitude. Although it might appear that more quantization noise is introduced in the process, it turns out that this scheme offers a clear advantage when the input signal swings outside the quantizer dynamic range. The details are deferred until Section \ref{sec:SQNR}.

\subsection{Sampling Rate Considerations}
Since UDR-ADC performs a modulo operation on the input signal samples $x(nT)$, $T$ being the sampling period, there is an inherent nonuniqueness in the input-output mapping, i.e., the output samples given by $y\left(nT\right)=\mathcal{M}_{V_{ref}} \{x\left(nT\right)+m2V_{ref}\}$ are the same for all $m \in {\mathbb{Z}^+}$. In order to avoid this scenario, we arrive at a sufficient condition on the sampling period $T$ such that the measurements do not change by more than $2V_{ref}$ in a sampling interval. Let us assume that the input signal $x(t)$ is Lipschitz-continuous, i.e., $x(t)$ satisfies the property: $|x(a)-x(b)|\leq \alpha |b-a|$, for some $\alpha \in \mathbb{R}^+$ and $\forall a,b \in \mathbb{R}$; the parameter $\alpha$ is the Lipschitz constant. Effectively, this property places a restriction on the growth-rate of the signal. The sufficient condition is obtained by enforcing that 
\begin{IEEEeqnarray} {RCL}
|x\left((n+1)T\right)-x(nT)|\leq \alpha T &&\leq 2\,V_{ref},\,\, \forall \,\, n, \nonumber\\
\Rightarrow T &&\leq \displaystyle\frac{2\,V_{ref}}{\alpha}.
\label{eq:sampling_interval}
\end{IEEEeqnarray} 
We would like to emphasize that there is no constraint of bandlimitedness in the preceding calculation. However, for the specific case of bandlimited signals also, one could use a similar growth-rate property \cite{Papoulis} and derive a sufficient condition on $T$.

\begin{figure*}[t]
 \centering
    \includegraphics[width=\textwidth]{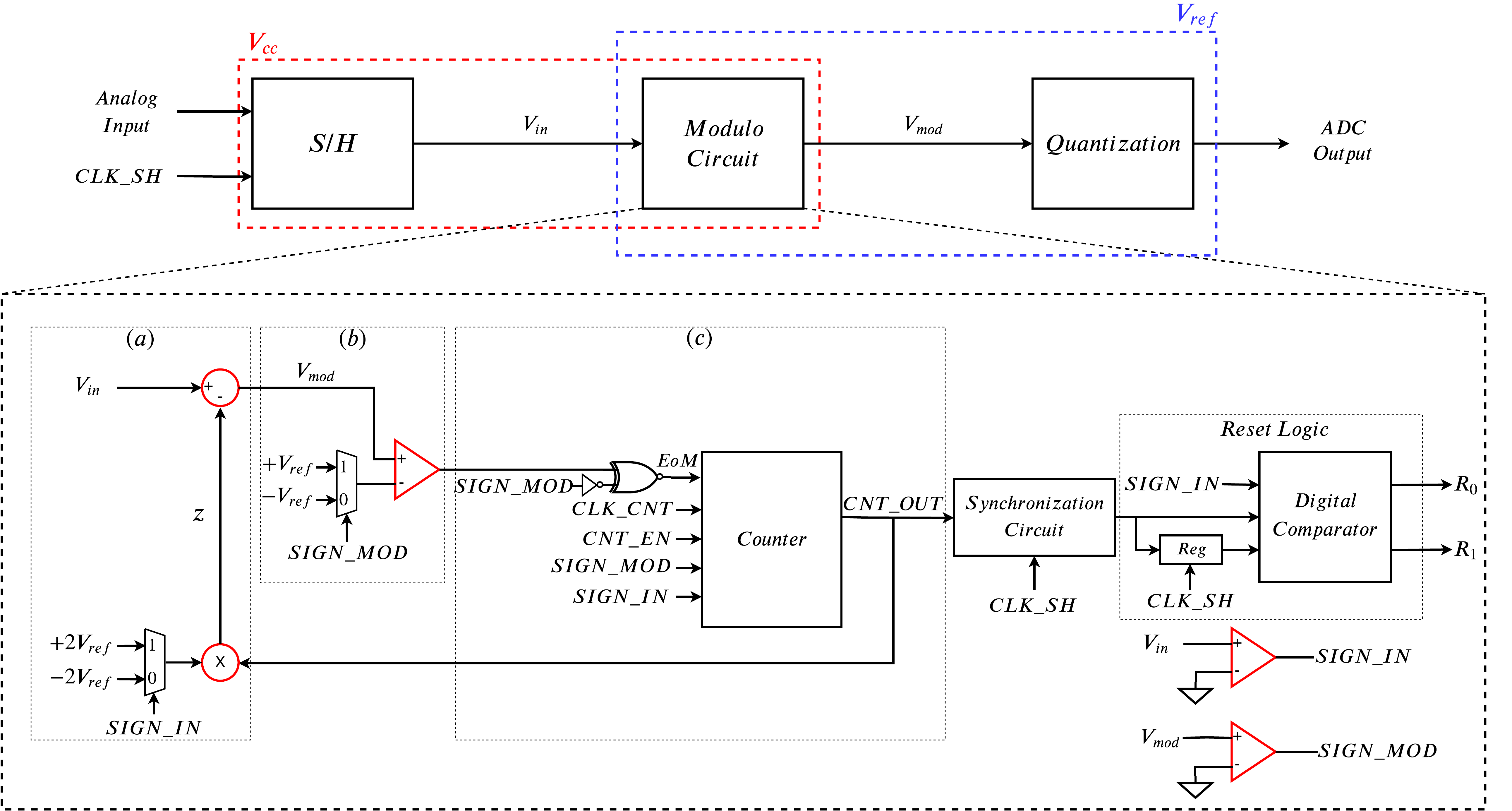}
\caption{(Color online) A system-level block diagram of the proposed UDR-ADC and the details of the modulo circuit. The modules highlighted in red operate at $V_{cc}$, which could be as high as allowed by the process technology, whereas the other blocks operate at $V_{ref} < V_{cc}$.}
\label{fig:mod_circuit}
\end{figure*} 
\begin{table*}[t]
\caption{Truth-table for the counter and reset logic. $\Delta\,CNT\_OUT$ indicates the change in $CNT\_OUT$ and $\Delta\,z$ denotes the change in $z$.}
\centering
\begin{tabular}{P{0.06\linewidth}|P{0.12\linewidth}|P{0.12\linewidth}|P{0.12\linewidth}|P{0.12\linewidth}|P{0.08\linewidth}|P{0.08\linewidth}}
\hline \hline
\text{\small {$EoM$}} & \text{\small {$SIGN\_MOD$}}  & \text{\small {$SIGN\_IN$}} & \text{\small {$\Delta\,CNT\_OUT$}} & \text{\small {$\Delta z$}} & Reset  & $R_1R_0$\\
 \hline \hline
 $0$ & $1$  & $1$ & $+1$ & $+2V_{ref}$ & Positive & $01$   \\ \hline
 $0$ & $0$  & $1$ & $-1$ & $-2V_{ref}$  & Negative & $11$\\ \hline
 $0$ & $1$  & $0$ & $-1$ & $+2V_{ref}$  & Positive & $01$\\ \hline
 $0$ & $0$  & $0$ & $+1$ & $-2V_{ref}$ & Negative & $11$\\ \hline
 $1$ & $\times$  & $\times$ & 0  & 0 & No reset & $00$ \\ \hline \hline
\end{tabular}
\label{tab:TT}
\end{table*}
\section{Unlimited Dynamic Range ADC Architecture}
\label{sec:architectures}
As alluded to in Section~\ref{sec:contribution}, the UDR-ADC consists of three blocks: (i) sample-and-hold circuit, (ii) modulo circuit, and (iii) quantizer. In this section, we present the operational details of the various blocks.

\subsection{The Modulo Circuit}
\indent The modulo circuit (cf. Fig.~\ref{fig:mod_circuit}) performs a modulo on the input $V_{in}$ and results in the output $V_{mod}=\mathcal{M}_{V_{ref}}\{V_{in}\}$ as per \eqref{eq:mod_op}. With respect to Fig.~\ref{fig:recon_demo}, $V_{in} = x[n]$, $V_{mod} = y[n]$. The input and output of the modulo circuit are sampled analog signals, i.e., real-valued signals defined in discrete-time. The modulo is performed with respect to $2V_{ref}$, which is either added to or subtracted from $V_{in}$, depending on the sign of $V_{in}$, and is performed by sub-block (a) in Fig.~\ref{fig:mod_circuit}. An opamp (sub-block (b)) compares the result $V_{mod}$ with the reference voltage $V_{ref}$ to determine whether it is within the range $[-V_{ref}, V_{ref}]$ or not. The logic to enable or disable addition or subtraction of $2V_{ref}$ to $V_{in}$ is performed by sub-block (c). The modulo circuit utilizes a counter and a feedback mechanism to perform these operations. The counter operates at a higher clock frequency ($CLK\_CNT$) than the sample-and-hold circuit ($CLK\_SH$) and keeps track of the number of times $2V_{ref}$ has been added to or subtracted from $V_{in}$. To start with, the counter output $CNT\_OUT$ is set to zero and is enabled by making $CNT\_EN$ high. The signals $SIGN\_IN$ and $SIGN\_MOD$ represent the polarities of $V_{in}$ and $V_{mod}$, respectively. For instance, if $V_{in}$ is positive, $SIGN\_IN$ will be high, otherwise it will be low.  As the counter output is unsigned, during feedback, $CNT\_OUT$ is multiplied with $\pm 2V_{ref}$ based on $SIGN\_IN$.\\
\indent The input $V_{in}$ obtained using a sample-and-hold circuit is fed to the subtractor and its output ($V_{mod}$) is compared with the reference voltage $(\pm V_{ref})$ using an opamp comparator. If $|V_{mod}|<|V_{ref}|$, it indicates the end of modulo ($EoM$) operation, and $EoM$ goes high keeping the counter output unchanged. On the other hand, if $|V_{mod}|>|V_{ref}|$, the $EoM$ signal will be low and the $CNT\_OUT$ is either incremented or decremented based on the signs of $V_{mod}$ and $V_{in}$ according to the truth-table given in Table~\ref{tab:TT}. \\
\indent Based on the sampling rate consideration, $CNT\_OUT$ may change by $\pm 1$ or remain unchanged compared with its previous value. Thus, the modulo circuit takes a maximum of two cycles of $CLK\_CNT$ in order to generate the result $V_{mod}$. After the modulo is completed, $CNT\_EN$ goes low, and $QNT\_EN$ goes high in order to activate the quantization block. The sample-and-hold circuit should hold the input sample $V_{in}$ for at least two cycles of the counter clock $CLK\_CNT$ and the delay $\tau$ introduced by the quantization block, i.e., $T_{CLK\_SH} \geq 2\,T_{CLK\_CNT} + \tau$, where $T_{\{\cdot\}}$ denotes the corresponding time period. The synchronization circuit ensures that the data transfer between sub-block (c) and the Reset logic block is synchronized.

\begin{figure*}[!t]
\centering
\includegraphics[width=14cm]{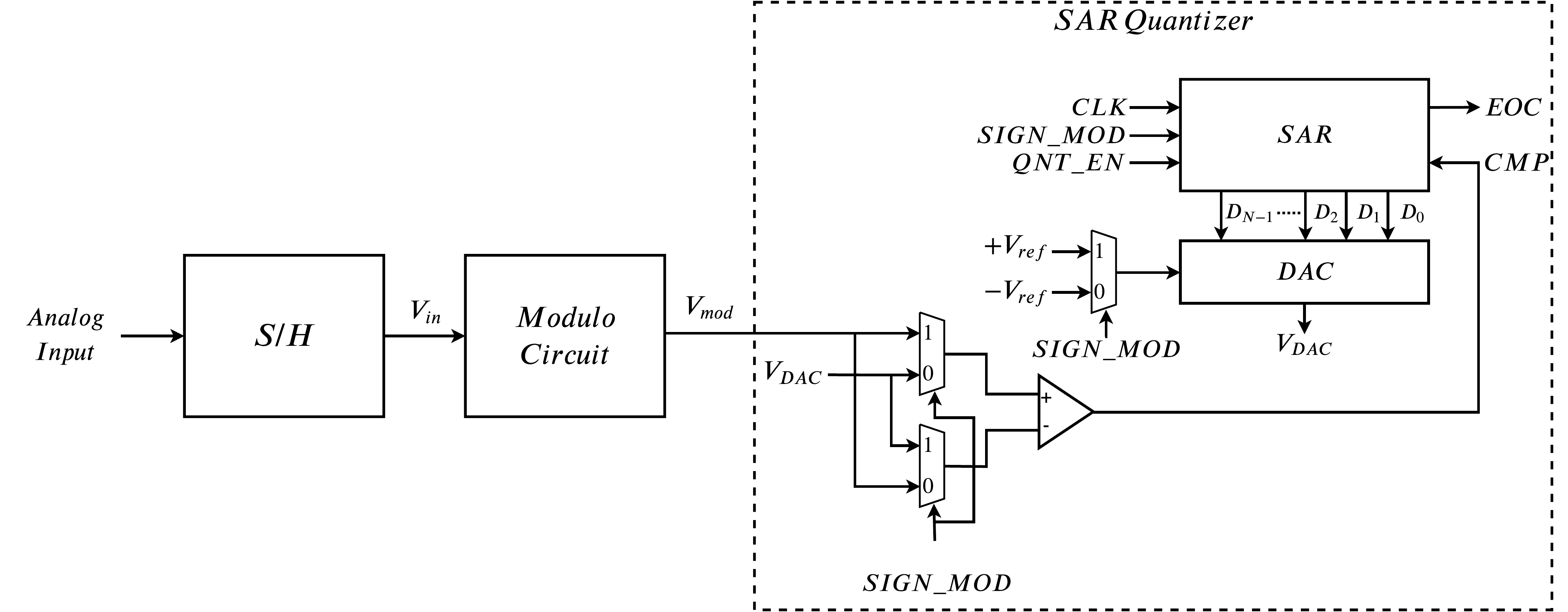}
\centering\caption{Schematic representation of a SAR UDR-ADC. The SAR quantization block operates on bipolar input voltages. The quantized bits of interest are $(D_{N-1}D_{N-2}\cdots D_0)$, $D_0$ being the LSB and $D_{N-1}$ being the MSB.}
\label{fig:SAR_ADC}
\end{figure*} 
\subsection{Operation of the Reset Logic Block}
As mentioned in Section~\ref{sec:reconstruction}, two bits $R_1R_0$ are employed to encode the three reset possibilities (positive reset, negative reset, or no reset). For every input sample, at the end of the modulo operation, the $CNT\_OUT$ value is compared with that of the previous sample and encoded using $R_1R_0$. The `reset' logic and the corresponding truth-table are shown in Fig.~\ref{fig:mod_circuit} and Table~\ref{tab:TT}, respectively. The column $\Delta z$ in the truth-table denotes the change in the signal value that is fed to the subtractor circuit in sub-block (a) of Fig.~\ref{fig:mod_circuit}, and corresponds to the change in $z[n]$ (cf. Fig.~\ref{fig:recon_demo}(c) for $z[n]$).\\
\indent Next, we present details of the quantization circuit employing the SAR module. Since the modulo circuit is generic, one could also employ the quantization blocks of flash or dual-slope ADCs instead of SAR.

\subsection{SAR UDR-ADC }
A SAR ADC employs binary search for discretizing an analog signal. It utilizes a comparator that successively compares its input $V_{mod}$ (which is the output of the modulo circuit) with the output of a digital-to-analog converter (DAC), thereby setting the registers appropriately and arriving at the corresponding quantization level. Figure~\ref{fig:SAR_ADC} shows a system-level block diagram of the proposed SAR UDR-ADC. The SAR has $N$ bits ($D_0,\, D_1,\, \cdots, D_{N-1}$) for quantization and a sign bit $(SIGN\_MOD)$ to indicate the polarity of the input. The proposed quantization circuit can deal with both positive and negative signal voltages unlike the conventional SAR model. We next analyze the operation of the quantizer based on the polarity of $V_{mod}$.

\subsubsection*{ \textbf{Case 1}} $V_{mod} \geq 0$\\
In this case, $SIGN\_MOD$ goes high and a DAC converts the SAR value to its analog counterpart $V_{DAC}$, which is compared with $V_{mod}$ (cf. Fig.~\ref{fig:SAR_ADC}). The SAR bits are set/reset appropriately to arrive at a suitable digital representation of $V_{mod}$ and at the end of the digitization operation, the end-of-conversion (EoC) signal goes high. The operation in this case is similar to that of a conventional SAR ADC.
\subsubsection*{ \textbf{Case 2}} $V_{mod} < 0$ \\
In this case, $SIGN\_MOD$ goes low and the reference voltage of the DAC is changed to $-V_{ref}$. Initially, the SAR is set to $(100 \cdots 00)$ and the DAC output $V_{DAC}$ is compared with $V_{mod}$. If $|V_{mod}| > |V_{DAC}|$, the MSB of SAR is retained as $1$, else it is set to $0$. The conversion operation as in Case 1 is repeated to get an equivalent $N$-bit representation of $|V_{mod} |$. At the end of the operation, the $EoC$ pulse goes high and the SAR output is inverted. To get the digital equivalent of $V_{mod}$, inside the SAR block, XOR operation is performed on each of the SAR output bits with that of $\overline{SIGN\_MOD}$ bit.

\section{Experimental Results}
\label{sec:results}
In this section, we present circuit simulations and a functional UDR-ADC realized using discrete components.
 \begin{figure}[t]
  \centering
    \includegraphics[width=3.5 in]{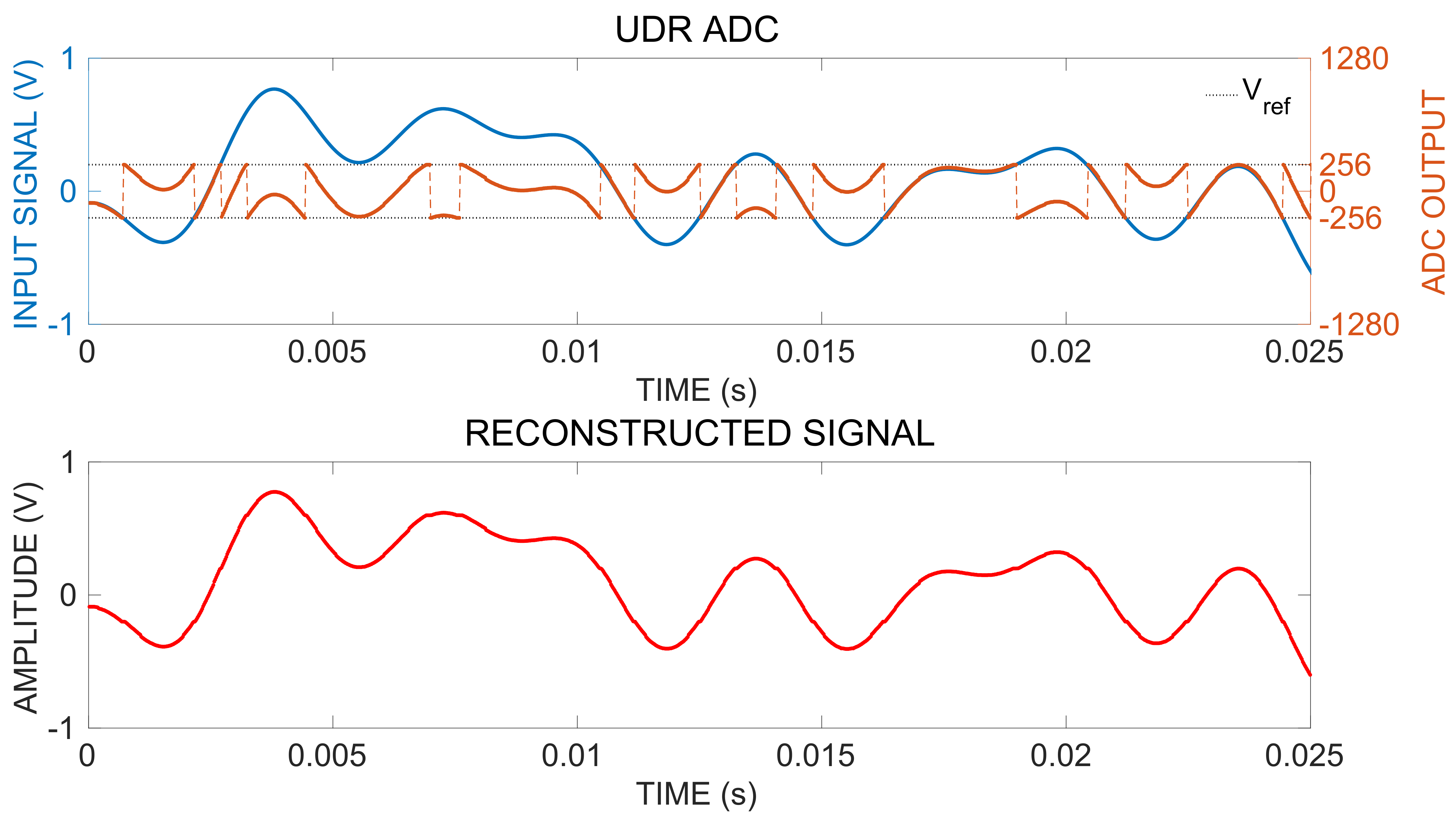}
  \caption{[Color online] CMOS circuit simulation results: The top plot shows the input analog signal (a sum of sinusoids) and the modulo samples of the UDR-ADC. The plot of the modulo samples must be read with the help of the $y$-axis shown on the right-hand side, which corresponds to a 9-bit representation. The bottom plot shows the signal reconstructed from the UDR-ADC measurements.}
  \label{ADC_results}
\end{figure}
 \begin{figure*}[t]
  \centering
    \begin{tabular}{ll}
  \hspace{-0.2 in} \raisebox{-0.5\height}{ \includegraphics[width=3.3 in]{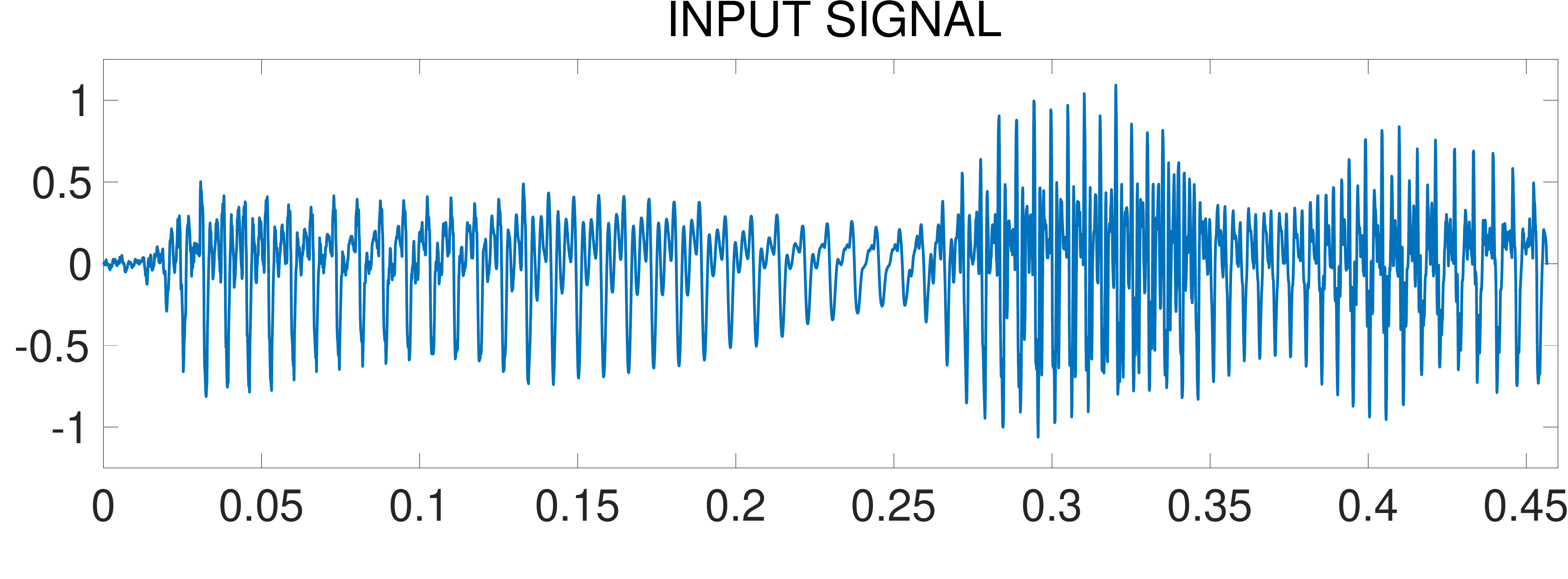}} \hspace{-0.03 in}{(a)} \quad \quad  \hspace{-0.2 in} \raisebox{-0.5\height}{ \includegraphics[width=3.3 in]{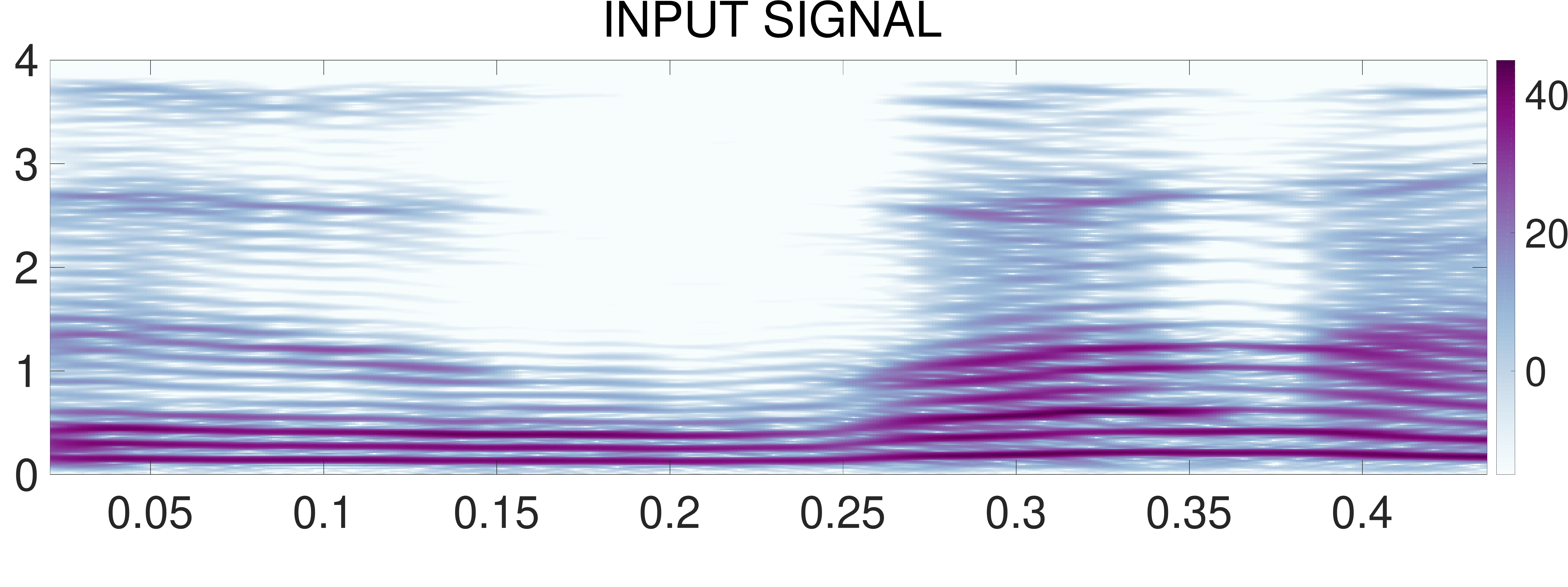}} \hspace{-0.03 in}{(d)}\\
  \hspace{-0.2 in} \raisebox{-0.5\height}{ \includegraphics[width=3.3 in]{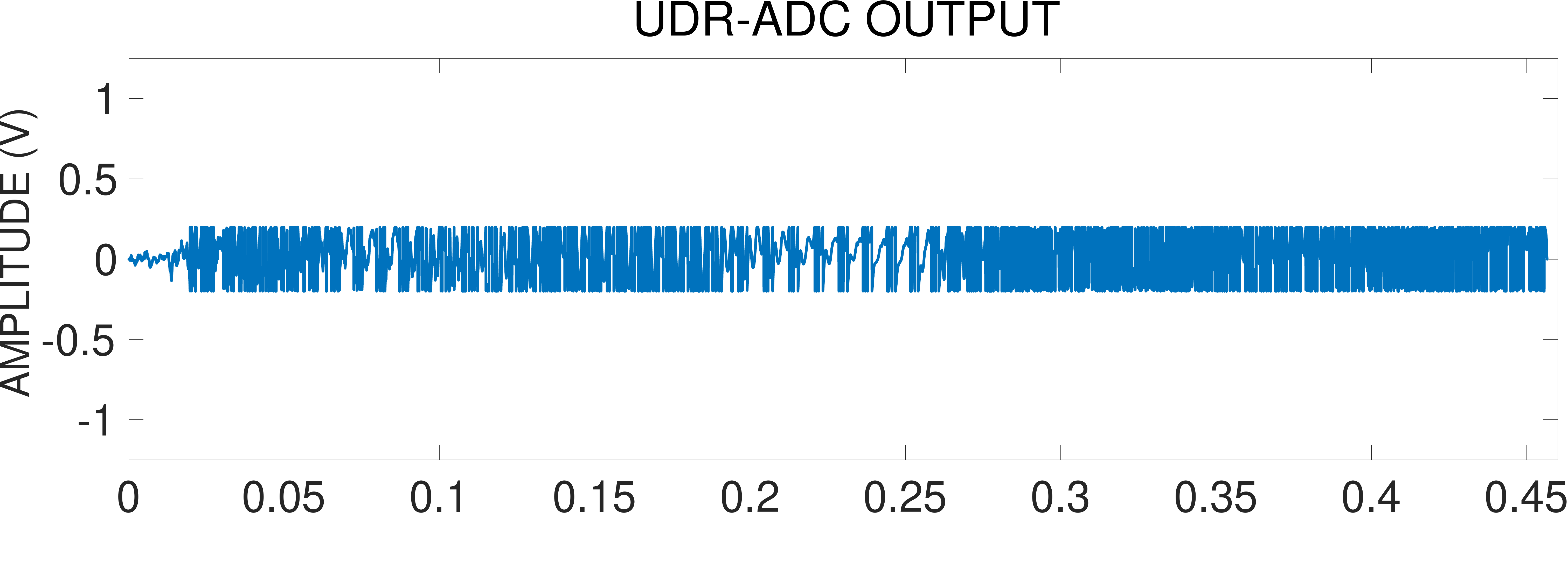}} \hspace{-0.03 in}{(b)} \quad \quad \hspace{-0.2 in} \raisebox{-0.5\height}{  \includegraphics[width=3.3 in]{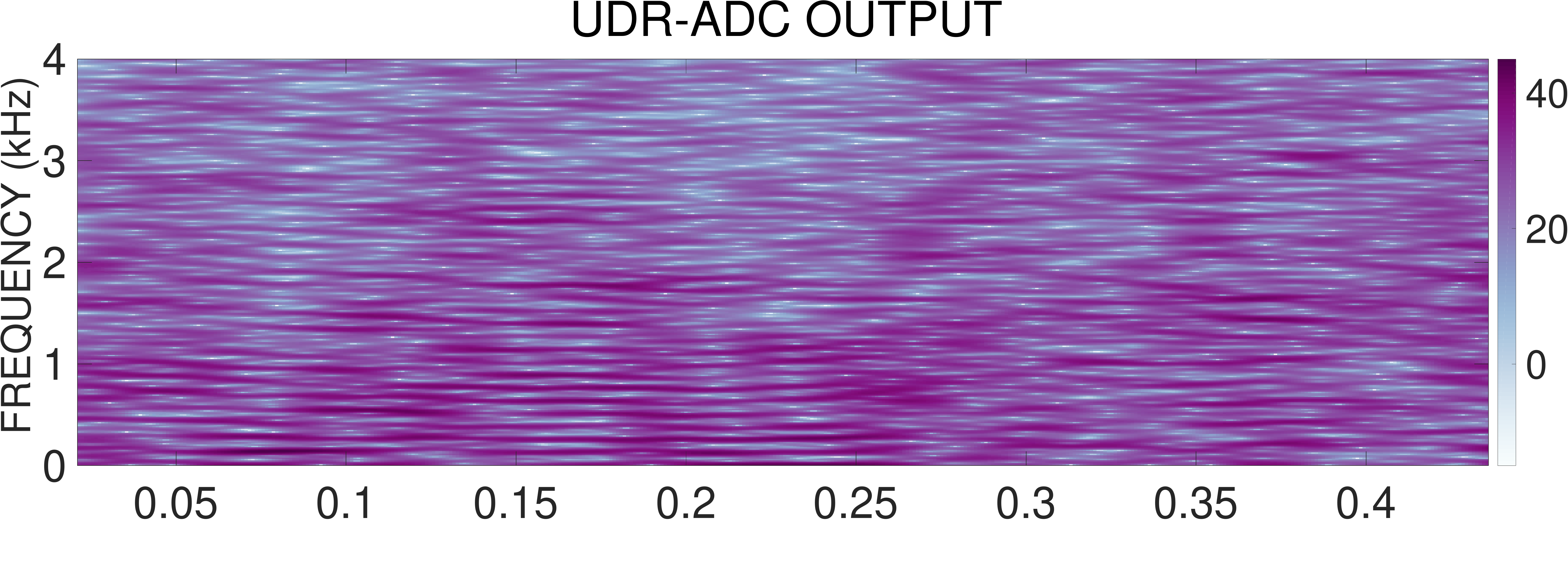}} \hspace{-0.03 in}{(e)}\\
 \hspace{-0.2 in} \raisebox{-0.5\height}{ \includegraphics[width=3.3 in]{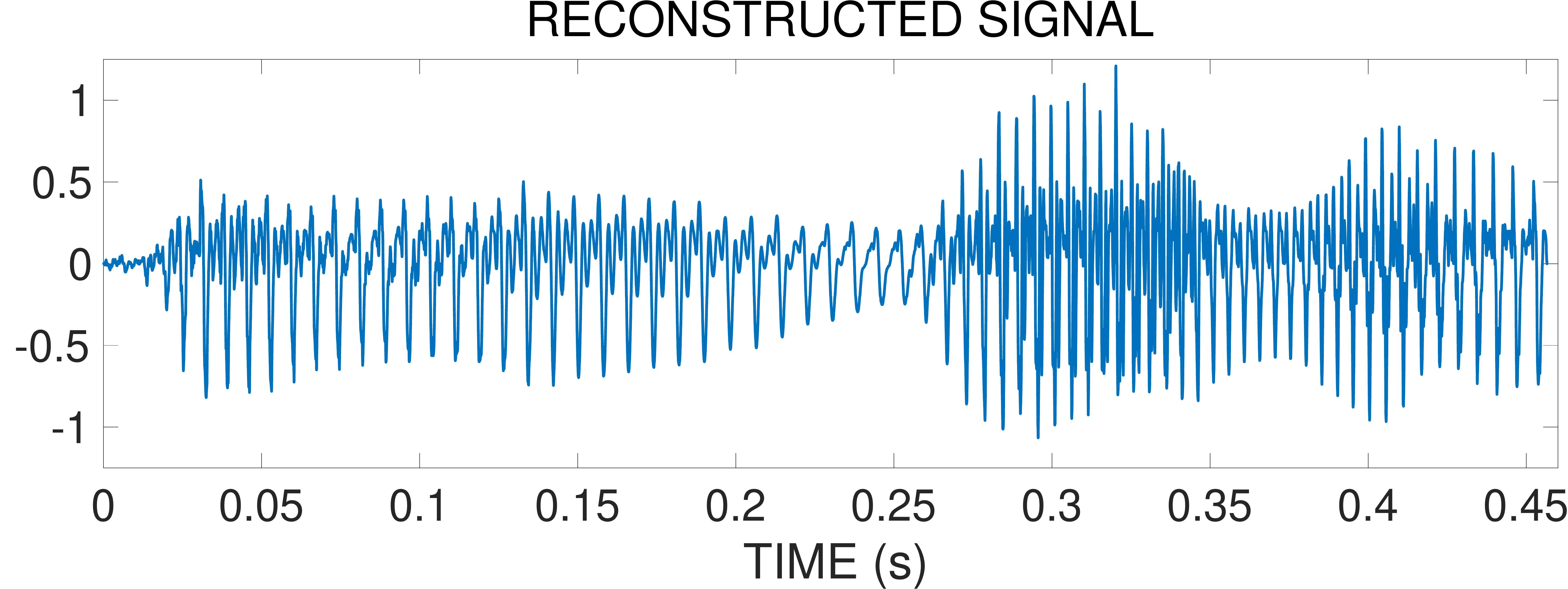}} \hspace{-0.03 in}{(c)} \quad \quad  \hspace{-0.2 in} \raisebox{-0.5\height}{  \includegraphics[width=3.3 in]{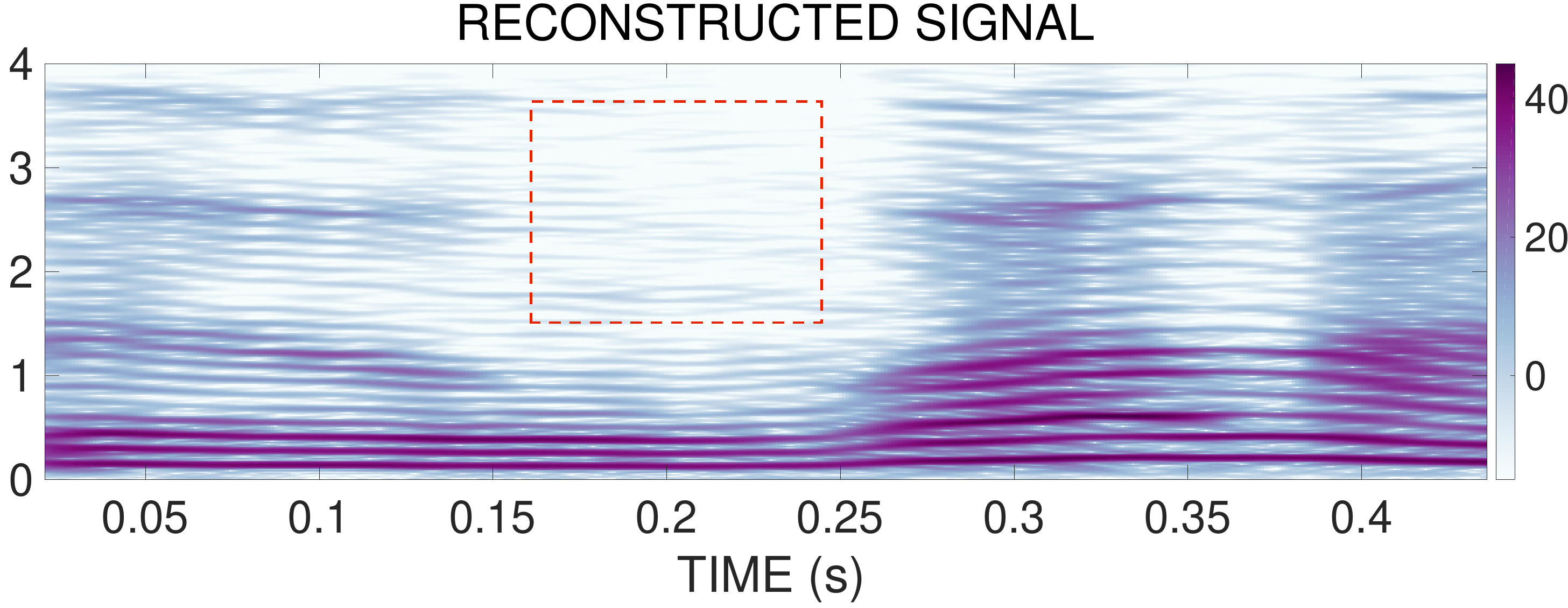}} \hspace{-0.03 in}{(f)}
   \end{tabular}
  \caption{[Color online] UDR-ADC CMOS circuit simulation results with a speech signal as the input: (a)-(c) Time-domain waveforms; and (d)-(f) the corresponding spectrograms. The signal-to-reconstruction-error ratio is 63.36 dB, indicating a high accuracy. The reconstruction error is largely due to quantization noise, which could be seen in the spectrogram of the reconstructed signal (dashed box).}
\label{fig:Spect}
 \end{figure*}
\subsection{Circuit Simulation}
\indent We implemented the SAR UDR-ADC in Cadence design environment with $65$ nm CMOS process technology. A double-buffered sample-and-hold circuit with CMOS switch was implemented as per the design by van de Plassche \cite{van2013cmos}. The subtractor and multiplier were implemented using opamps. A two-stage unity-gain Miller-compensated opamp is realized based on the design proposed by Allen and Holberg \cite{allen2010cmos}. Digital circuits such as SAR, counter, and reset logic are modeled in Verilog. A total of  $11$ bits per sample are allocated for A/D conversion, out of which, $9$ are used for quantization and $2$ for encoding the reset information. An R-2R binary ladder is used as the DAC. Multiplication of $CNT\_OUT$ with $\pm 2V_{ref}$ is achieved by multiplying $\pm V_{ref}$ with $CNT\_OUT$ value left-shifted by $1$ bit. This eliminates the need for a separate voltage source operating at $\pm 2V_{ref}$.\\
\indent To illustrate the performance of the circuit, we first consider a sum of sinusoids consisting of frequencies $70,\, 30,\, 200,$ and $300$ Hz. The maximum amplitude of the input is $1.2$ V and the reference voltage of the UDR-ADC is set to $0.2$ V.  The operating frequency of the sample-and-hold circuit is chosen as $53$ kHz. Figure~\ref{ADC_results} shows the input signal, the output samples of the UDR-ADC, and the reconstructed signal. The figure shows that the proposed circuit is capable of successfully implementing the modulo operation. The reconstruction from the modulo measurements and the reset information is also accurate. The accuracy is quantified by comparing  the reconstruction against the input signal sampled at 53 kHz and quantized using 32 bits per sample. The signal-to-reconstruction-error ratio was computed to be 75.24 dB, which indicates a high accuracy.\\

\indent The next illustration employs a speech signal of bandwidth 4 kHz as the input. The reference voltage of the ADC is set to $0.2$ V, whereas the maximum voltage of the  input signal is $1.2$ V. Figures~\ref{fig:Spect}(a)-(c) show the input speech signal, the modulo samples, and the reconstructed signal, respectively. The corresponding spectrograms are shown in Figures~\ref{fig:Spect}(d)-(f). The signal-to-reconstruction-error ratio in this case turned out to be 63.36 dB.

\subsection{Hardware Prototype}
\begin{figure*}[!t]
  \centering
  \begin{subfigure}[b]{0.68\textwidth}
    \includegraphics[width=\textwidth]{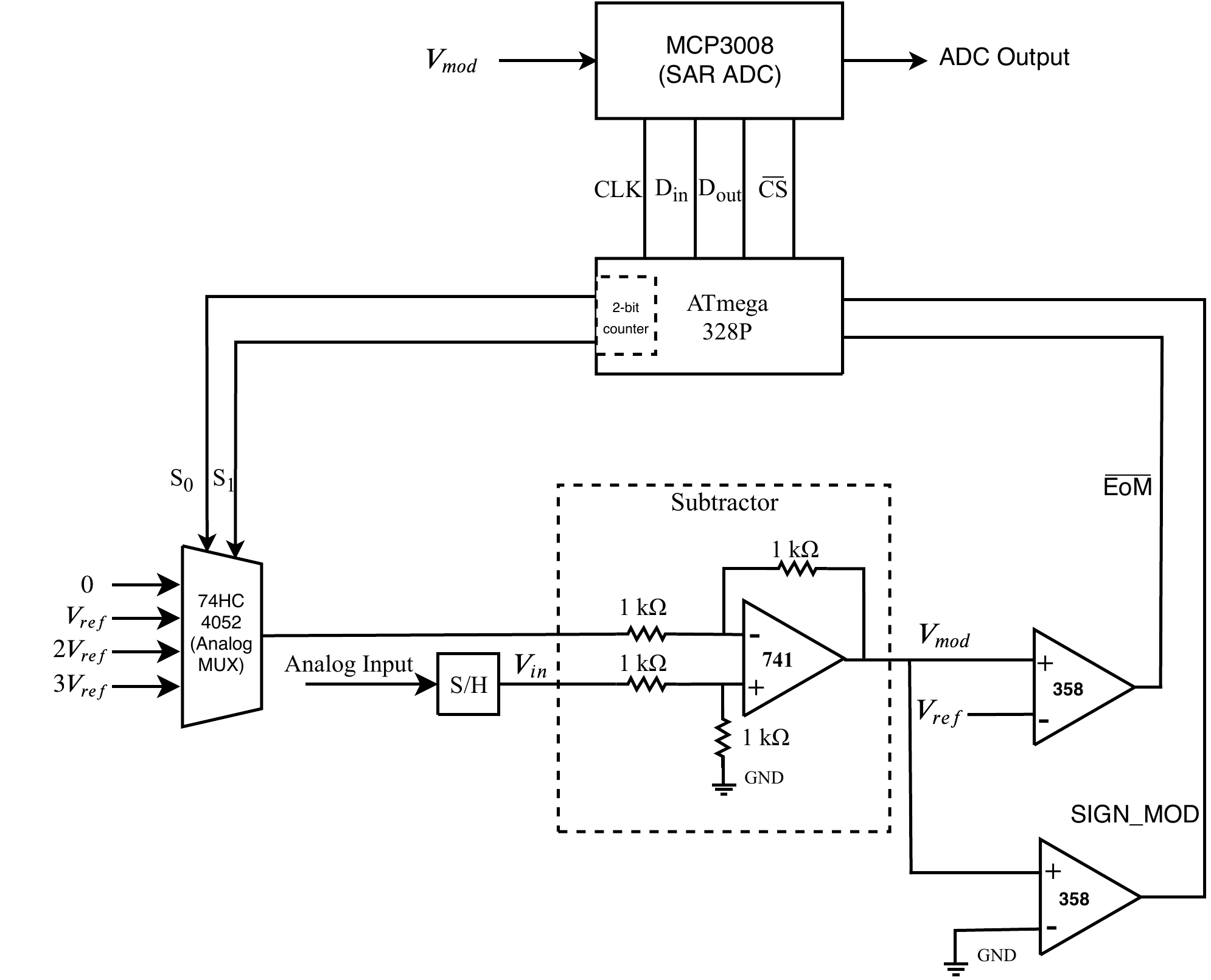}
    \caption{ }
  \end{subfigure}
  \begin{subfigure}[b]{0.45\textwidth}
    \includegraphics[width=\textwidth]{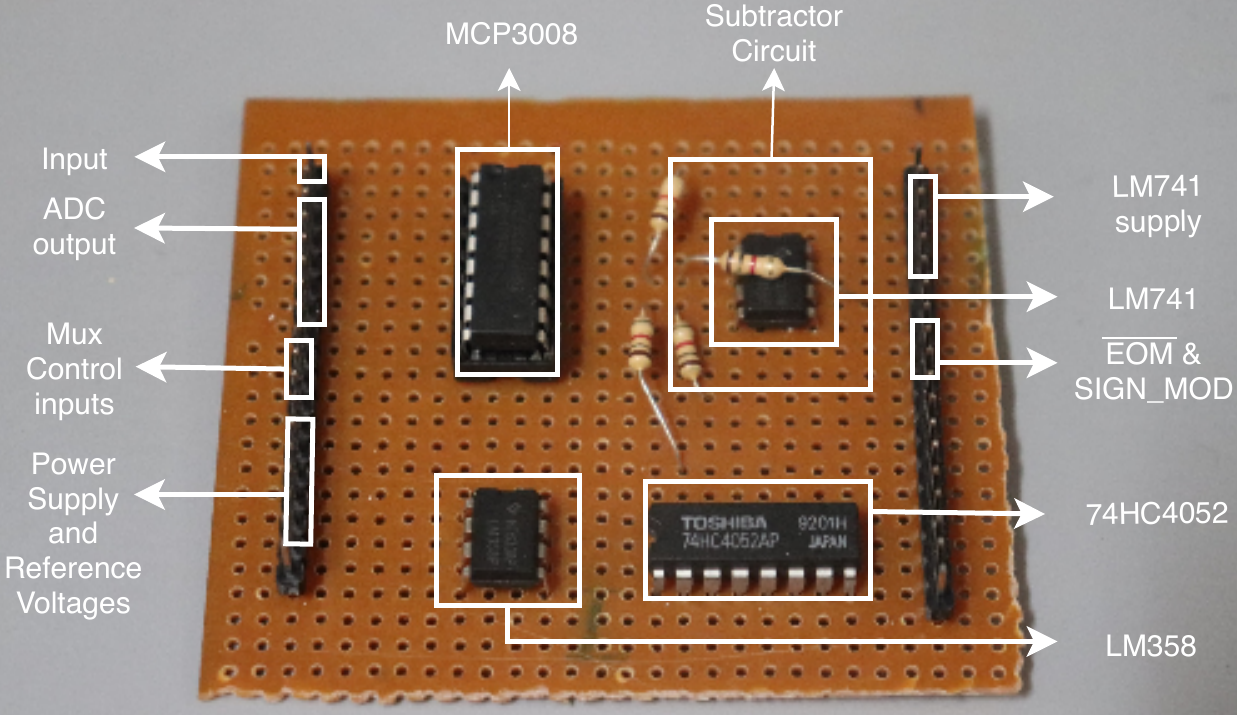}
    \caption{ }
  \end{subfigure}
  \begin{subfigure}[b]{0.5\textwidth}
    \includegraphics[width=\textwidth]{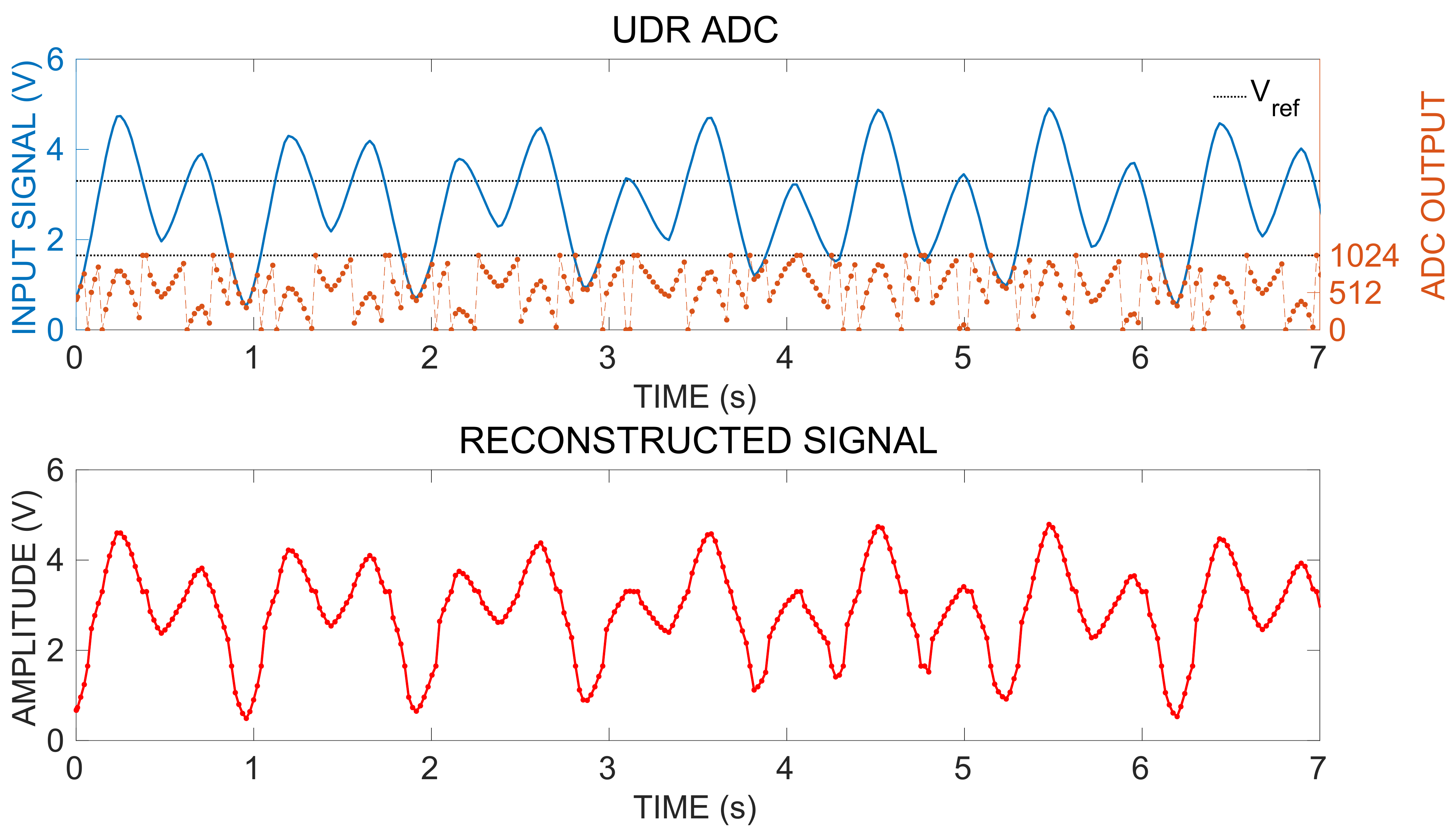}
    \caption{}
  \end{subfigure}
  \caption{Hardware prototype of a $12$-bit UDR-ADC demonstrating real-time implementation: (a) Block diagram; (b) A picture of the circuit board; and (c) Results pertaining to a sum-of-sinusoids input. The peak input amplitude is chosen as $5$ V and the reference voltage $V_{ref}=1.65$ V, which results in a maximum of three resets per sample.}
  \label{ADC_prototype}
\end{figure*}

\indent The block diagram that forms the basis for our prototype is shown in Fig.~\ref{ADC_prototype}(a). We employed a standard SAR-ADC integrated circuit (IC) MCP3008 from Microchip Technology as it has the flexibility to choose the reference voltage as high as the power-rails. Since MCP3008 can handle only positive-valued inputs, the prototype is designed to handle only such inputs. The SAR-ADC is interfaced with the rest of the circuit using an ATmega328P microcontroller by serial peripheral interface (SPI) protocol and is programmed to work at $200$ kilo samples per second. The subtractor circuit is implemented using Texas Instrument's standard LM741 opamp and the offset voltages of the IC are adjusted by a trimming potentiometer provided in the IC. The circuit is built to accommodate up to three resets per sample (i.e., a folding factor of $3$). To achieve the reset operation, a $4\times2$ analog multiplexer IC 74HC4052  is employed (cf. Fig.~\ref{ADC_prototype}(a)). A single power-rail IC LM328 is used as the comparator for the end of modulo ($EoM$) as well as for the polarity of $V_{mod}$ ($SIGN\_MOD$). Based on $EoM$ and $SIGN\_MOD$, the counter implemented in ATmega328P is either incremented or decremented appropriately. The control bits $S_0$ and $S_1$ for the multiplexer are provided depending on the counter output. The control bits also act as the reset bits. The prototype UDR-ADC is shown in Fig.~\ref{ADC_prototype}(b). For illustration, a signal consisting of a mixture of sinusoids is given as the input to the prototype. Figure~\ref{ADC_prototype}(c) shows the input signal (measured by a digital storage oscilloscope), the corresponding modulo samples (output of the UDR-ADC), and the reconstructed signal (in MATLAB). The hardware implementation results reaffirm the inferences made from circuit simulations.
\section{Performance Analysis}
\label{sec:performance}
In this section, we compare the performance of the UDR-ADC vis-\`a-vis a standard ADC in terms of the area and power requirements, and signal-to-quantization-noise ratio (SQNR). Denote the {\it folding factor} as $\lambda = V_{max}/V_{ref}$, where $V_{max}$ is the maximum amplitude of the input analog signal and $V_{ref}$ is the reference voltage of the UDR-ADC with respect to which the modulo happens. For instance,  
$\lambda = 4$ indicates that the input signal amplitude is more than $4V_{ref}$.
\subsection{Area}
\label{sec:Area}
\begin{figure}[t]
\centering
\includegraphics[width=3.5 in]{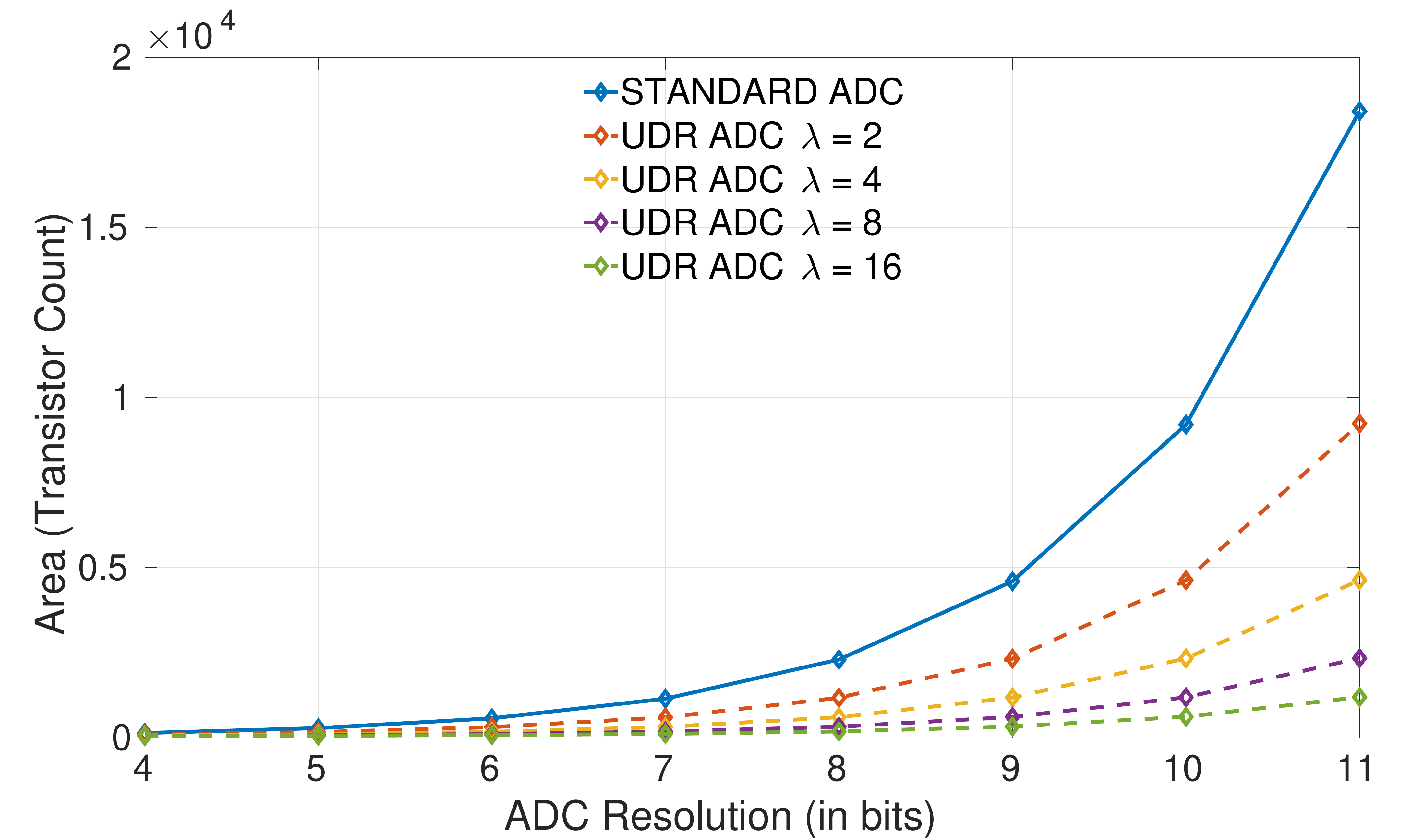}
\caption{[Color online] A comparison of the area required for the standard flash ADC and the flash UDR-ADC for different folding factors ($\lambda$) as a function of the number of bits. The UDR-ADC has a significantly lower area than the standard ADC.}
\label{fig:area_analysis}
\end{figure}

The area required by an ADC is measured in terms of the transistor count, which in turn depends on the number of bits used for digitization. Let $n_1$ be the number of bits employed in a UDR-ADC, corresponding to which the quantization step is given by
\begin{equation}
\Delta_{UDR}=\frac{2V_{ref}}{2^{n_{1}}}.
\label{eqn:delta_UDR}
\end{equation}
Let $n_2$ be the number of bits employed in a flash ADC corresponding to the same quantization step size, i.e.,
\begin{align}
\Delta_{STD}&=\frac{2V_{max}}{2^{n_{2}}} =\Delta_{UDR}=\frac{2V_{ref}}{2^{n_{1}}}, \nonumber \\
2^{n_2-n_1}& = \frac{V_{max}}{V_{ref}} =\lambda.
\label{eqn:delta_STD}
\end{align}
Considering $V_{max} \geq V_{ref}$, the number of bits required by a flash ADC is given by 
\begin{equation}
    n_{2} = n_{1} + \lceil \log_{2} \lambda \rceil, 
    \label{eqn:Bit_size}
\end{equation}
which is $\lceil \log_{2} \lambda \rceil$ bits more than that of a UDR-ADC for the same quantization step-size.\\
\indent In a standard flash ADC, the area increases exponentially with the number of bits as it requires $2^{n_{2}-1}$ comparators and $2^{n_{2}}$ resistors. On the other hand, a UDR-ADC requires less number of bits ${n_{1}}$ (cf. \eqref{eqn:Bit_size}), which in turn reduces the number of comparators and resistors required. Also, with an increase in the folding factor $\lambda$, the area required by the UDR-ADC further reduces as shown in Fig.~\ref{fig:area_analysis} for various folding factors. As $\lambda$ increases, the area utilized by the additional circuitry required to perform the modulo operation also increases, but the increase is small compared to the total area.

\subsection{Power}
\label{sec:power}
\begin{figure}[t]
\centering
\includegraphics[width=3.5 in]{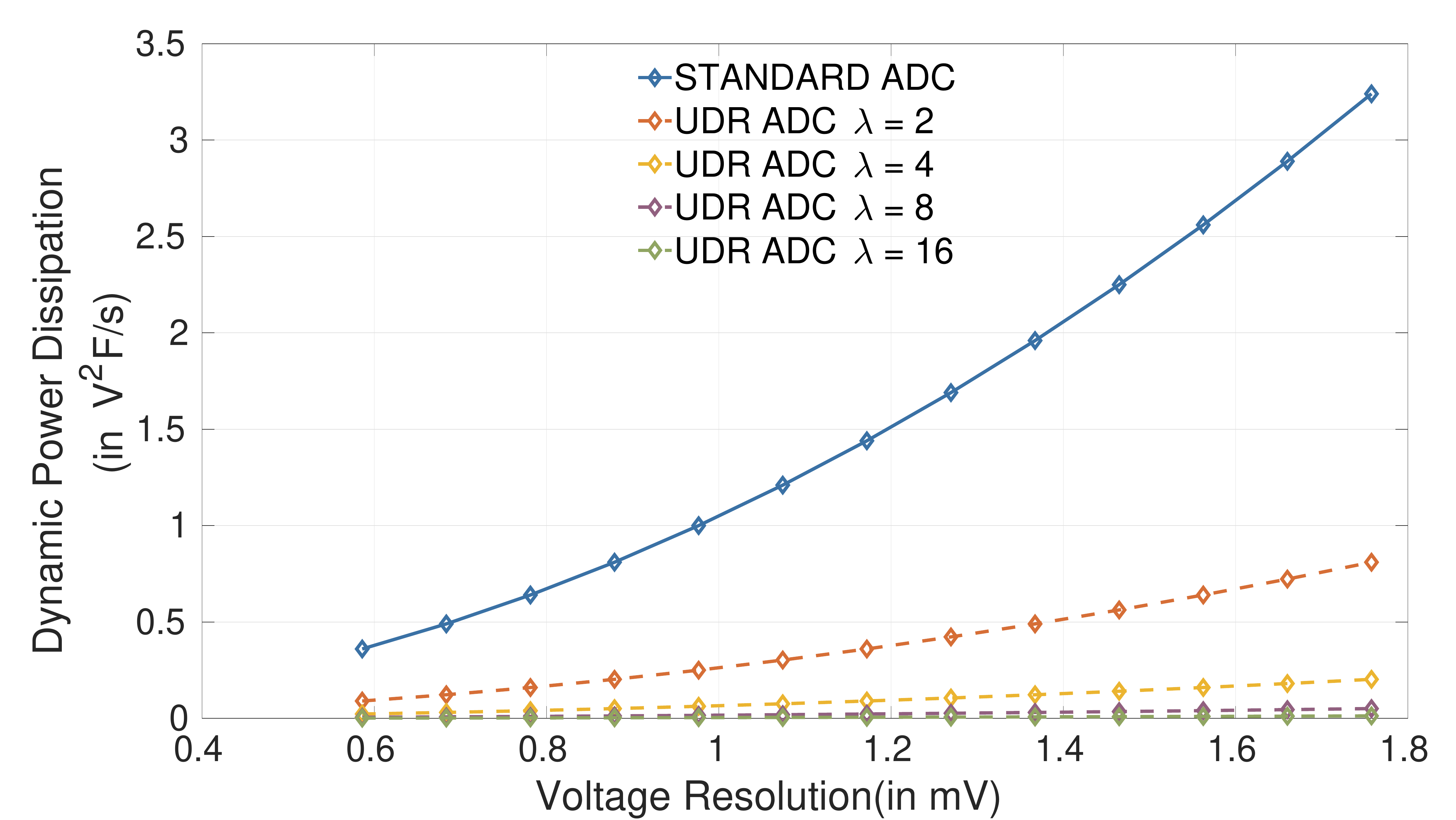}
\caption{[Color online] A comparison of the dynamic power dissipation in a standard ADC versus UDR-ADC for different folding factors ($\lambda$) as a function of the voltage resolution or equivalently, the quantization step-size.}
\label{fig:power_analysis}
\end{figure}

Since the sample-and-hold circuit is common to both ADCs, we restrict the comparison to the quantization block. The dynamic power dissipation $P_{d(STD)}$ in an ADC is proportional to the supply voltage $V_{s(STD)}$ and the operating clock frequency $f_{clk(STD)}$ as given by
\begin{equation}
   P_{d(STD)} \propto V_{s(STD)}^2f_{clk(STD)}.
    \label{eqn:power_eqn}
\end{equation}
For a given quantization step-size, the reference voltage of a UDR-ADC ($V_{ref}$) could be made lower than the threshold of the standard ADC. Hence, the voltage supply $V_{s(UDR)}$ to the quantization block of a UDR-ADC could also be made lower than that of the standard ADC by the folding factor $\lambda$, i.e.,
\begin{equation}
    V_{s(UDR)} = \frac{V_{s(STD)}}{\lambda}.
    \label{eqn:voltage_eqn}
\end{equation}
Consequently, the overall dynamic power dissipation in the quantizer of a UDR-ADC is given by 
\begin{align}
    P_{d(UDR)} &\propto V_{s(UDR)}^2 f_{clk(UDR)} = \frac{V_{s(STD)}^2 f_{clk(STD)}}{\lambda^2}, \nonumber \\
    P_{d(UDR)}& = \frac{P_{d(STD)}}{\lambda^2},
    \label{eqn:oversampling}
\end{align}
which is $\lambda^2$ times lesser than that of standard ADC. The price to pay for the reduced power of the quantizer is the power dissipation in the modulo circuit in a UDR-ADC.\\
\indent Figure~\ref{fig:power_analysis} shows plots of the dynamic power dissipated for a unit capacitance. The dynamic power dissipation of the standard ADC is higher than that of UDR-ADC for a given quantization step-size. As the folding factor increases, the dynamic power dissipation of the UDR-ADC decreases.\\
\indent The static power dissipation is directly proportional to the transistor count. Since the area required in a UDR-ADC is smaller than that of the standard ADC (as discussed in Section \ref{sec:Area}), the static power dissipation is lower in a UDR-ADC.
\begin{table*}[t]
\caption{SQNR for the standard and UDR-ADCs for three input distributions. In the case of the standard ADC, all $n$ bits are used for quantization, whereas in the case of the UDR-ADC, $n-2$ bits are used for quantization and the remaining $2$ bits for encoding the reset information. The loading factor $\gamma = \frac{V_{ref}}{\sigma_x}$, where $\sigma_x$ is the standard deviation of the input.}
\centering
\begin{tabular}{M{0.11\textwidth}|M{0.28\textwidth}|M{0.27\textwidth}|M{0.23\textwidth}} 
\hline \hline
 SQNR & Uniform distribution & Gaussian distribution & Laplacian distribution \\ \hline \hline \\ [-1em]
Standard ADC& $\displaystyle\frac{1}{\frac{\gamma^2}{3\,( 2^{2n})}+\left(1-\frac{\gamma}{\sqrt{3}}\right)^3\,\mathbf{1}_{\gamma \in [0, \sqrt{3}]}}$  &  
$\displaystyle\frac{1}{\displaystyle\frac{\gamma^2}{3\,( 2^{2n})}+ \psi(\gamma)} $ \hspace{2.3cm}{cf. \eqref{eq:ov_std_gau} of Sec.~\ref{sec:SQNR_der_gau} for $\psi(\cdot)$}& 
$\displaystyle\frac{1}{\displaystyle\frac{\gamma^2}{3\,( 2^{2n})}+ e^{-\sqrt{2}\gamma}}$  \\ \\ [-1em]\hline \\ [-1em] 
UDR-ADC &  \multicolumn{3}{c} {$\displaystyle\frac{3\,( 2^{2n})}{16\gamma^2}$}  \\   [-1em] \\
\hline \hline 
\end{tabular}
\label{tab:SQNR}
\end{table*}
\begin{figure*}[t]
\centering
\includegraphics[width=7 in]{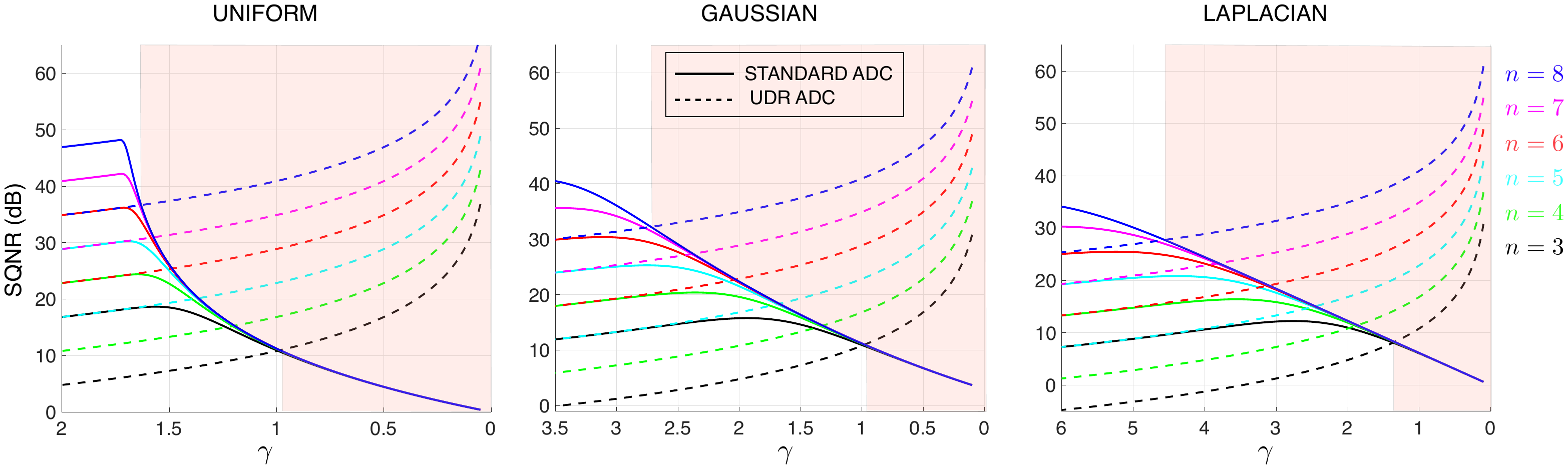}
\caption{[Color online] SQNR versus loading factor $\gamma$ for various input distributions. The number of bits ($n$) used for quantization are color-coded and indicated at the extreme right of the figure. The solid lines correspond to the standard ADC and the dashed lines correspond to the UDR-ADC. The shaded areas in the figure panels highlight the region in which the UDR-ADC outperforms the standard ADC.} 
\label{fig:SQNR}
\end{figure*}
\subsection{Signal-to-Quantization Noise Ratio}
\label{sec:SQNR}
Signal-to-quantization-noise ratio is an important parameter used to evaluate the performance of an ADC. It is defined as $$\text{SQNR} = \frac{\text{Signal variance}}{\text{Quantization noise variance}\,+\,\text{Overload distortion}}.$$ The quantization noise is a function of the threshold of the ADC and the number of bits used for digitization. A standard ADC introduces quantization noise and also overload distortion when the input signal amplitude exceeds $V_{ref}$, whereas the UDR-ADC introduces only quantization noise. But then, the UDR-ADC requires two bits for encoding the reset information. Hence, the SQNRs are not the same and depend on the parameters $V_{ref},\,V_{max},\, n$, and the distribution of the input signal amplitudes. In this work, we consider three types of input distributions: uniform, Gaussian, and Laplacian, and confine the analysis to the uniform quantization scheme. Table~\ref{tab:SQNR} summarizes the SQNRs for the various input distributions. The derivations of the SQNR formulae are provided in Appendix~\ref{sec:SQNR_derivation}. Figure~\ref{fig:SQNR} depicts the SQNR trends versus the {\it loading factor} $\gamma$ for various distributions. The loading factor is a measure of how much in excess of $V_{ref}$ the input signal swings, and is defined as $\gamma = \frac{V_{ref}}{\sigma_x}$, where $\sigma_x$ is the standard deviation of the input distribution under consideration \cite{jayantnoll}. It may be noted that the loading factor axis is reversed, which is in keeping with the convention \cite{jayantnoll}.

\subsubsection{Uniform Distribution}
In this case, let the dynamic range of the input signal be limited to $[-V_{max}, V_{max}]$, which corresponds to a standard deviation of $\frac{V_{max}}{\sqrt{3}}$. Whenever the loading factor $\gamma_{\mathcal{U}}^{} >\sqrt{3}$, there is no overload distortion as the dynamic range of the signal is well within the saturation threshold of the standard ADC. Hence, the SQNR depends only on the quantization noise variance. Since $n$ and $n-2$ bits are used for quantization in the standard and UDR-ADCs, respectively, the quantization noise is less in the case of the former. Consequently, the SQNR is higher for the standard ADC whenever $\gamma_{\mathcal{U}}^{}>\sqrt{3}$.\\
\indent Next, consider the case where $\gamma_{\mathcal{U}}^{} \leq \sqrt{3}$, i.e., there is both quantization noise and overload distortion in a standard ADC. The UDR-ADC does not suffer from overload distortion by design and is expected to have a higher SQNR. However, since the number of bits available for quantization is two less than that of a standard ADC, there is a transition region in which the quantization noise of UDR-ADC is more than the quantization noise and the overload distortion put together of a standard ADC. Beyond this region, the UDR-ADC offers dramatic SQNR gains over the standard ADC. The cross-over point is a function of the number of bits. The regions beyond the cross-over where the UDR-ADC can be gainfully deployed are highlighted in the figures. 

\subsubsection{Gaussian and Laplacian Distributions}
In several practical applications, the probability distribution of the input signal amplitudes is modeled as a Gaussian or a heavy-tailed distribution such as the Laplacian. For higher values of the loading factor, the overload distortion in the case of standard ADC is negligible and hence the SQNR of the standard ADC is better than that of the UDR-ADC. On the contrary, for large input variance, or equivalently, a high input dynamic range, the SQNR of the standard ADC deteriorates and drops below that of the UDR-ADC. The SQNR curves for different values of $n$ are shown in Fig.~\ref{fig:SQNR}. Again, the shaded areas are the operating regions where the UDR-ADC offers a clear advantage over the standard ADC.  One can also observe that, for a given loading factor, the SQNR gain of UDR-ADC over a standard ADC is more for the Laplacian distribution than the Gaussian. This is expected because the Laplacian has a heavier tail than the Gaussian, which results in a higher probability of a larger dynamic range.

\section{Conclusions and Outlook}
\label{sec:discussion}
We proposed a novel ADC with the self-reset feature, which allows for an unlimited dynamic range at the input. The self-reset happens by means of a modulo sampler, and a pair of dedicated bits that encode the reset information. Given the modulo samples and the reset information, the reconstruction is straightforward. As an illustration, we showed how a SAR-ADC could be converted to a UDR-ADC by introducing the modulo circuit between the sample-and-hold and the quantization blocks. The proposed architecture was simulated using $65$ nm CMOS technology in Cadence design environment. Simulation results showed that the quality of signal reconstruction from the modulo measurements is highly accurate. A hardware prototype built using discrete components further supported the feasibility of a real-time realization. A performance assessment in terms of the area, power, and SQNR showed that the proposed UDR-ADC has definitive advantages over the standard ADCs thus making it an ideal candidate for applications requiring a high dynamic range, low area, and low power.\\
\indent The UDR-ADC proposed in this paper employed a modulo circuit whose input and output are sampled analog signals, i.e., unquantized signal amplitudes defined in discrete-time. Hence, it was placed exactly in between the sample-and-hold and the quantization blocks. On the other hand, if one were to realize the modulo circuit in the continuous-time domain, it could simply precede an existing ADC thereby enabling ready conversion to a UDR-ADC.

\appendices
\renewcommand\thesubsection{\thesection.\arabic{subsection}}
\renewcommand\thesubsectiondis{\thesectiondis.\arabic{subsection}}
\renewcommand\thesubsubsection{\thesubsection.\roman{subsubsection}}
\renewcommand\thesubsubsectiondis{\thesubsectiondis.\roman{subsubsection}}
\section{Derivation of SQNR Formulae}
\label{sec:SQNR_derivation}
Let the input signal dynamic range be $[-V_{max},\, V_{max}]$. The saturation level of the standard ADC, which is also the self-reset threshold of the UDR-ADC, is denoted by $V_{ref}$, and the number of bits used for quantization in a standard ADC is denoted by $n$. In a UDR-ADC, since two bits are used for encoding the reset information, the number of bits available for quantization is $n-2$. Considering uniform quantization, the step-size is $\Delta_{STD}=\frac{2V_{ref}}{2^{n}}$ for the standard ADC and $\Delta_{UDR}=\frac{2V_{ref}}{2^{n-2}}=\frac{8V_{ref}}{2^{n}}$ for the UDR-ADC. The loading factor is defined as $\gamma = \frac{V_{ref}}{\sigma_x}$, where $\sigma_x$ is the standard deviation of the input distribution. Considering high-rate quantization, the quantization noise is modeled as a uniformly distributed random variable.

\subsection{SQNR for UDR-ADC}
For all the three signal distributions,  the quantization noise variance for the UDR-ADC is given by $\sigma_{q,UDR}^2 = \frac{\Delta_{UDR}^2}{12}=\frac{16V_{ref}^2}{3\,( 2^{2n})}$. As there is no overload distortion in UDR-ADC, the SQNR is given by 
\begin{IEEEeqnarray} {RCL}
SQNR_{UDR}&= & \frac{\sigma_{x}^2}{\sigma_{q,UDR}^2}= \frac{\sigma_x^2}{\frac{16V_{ref}^2}{3\,( 2^{2n})}}=\frac{3(2^{2n})}{16\gamma^2}.\nonumber 
\label{eq:SQNR_sr_uni}
\end{IEEEeqnarray} 

\subsection{SQNR for Standard ADC -- Uniform Input Distribution}
\label{sec:SQNR_der_uni}
Consider the uniform distribution $f(x)=\frac{1}{2V_{max}},\,x \in [-V_{max},\, V_{max}]$, and $0$ otherwise. The input variance is $\sigma_{x}^2 = V_{max}^2/3$. The quantization noise variance  is given by $$\sigma_{q,STD}^2 = \frac{\Delta_{STD}^2}{12}=\frac{V_{ref}^2}{3\,( 2^{2n})}$$ and the overload distortion is given by
\begin{IEEEeqnarray} {RCL}
\sigma_{ov,STD}^2 &= &2\int_{V_{ref}}^{V_{max}}(x-V_{ref})^2\,f_x(x)\,\mathrm{d}x, \nonumber \\
&=& \sigma_x^2\,\left(1-\frac{\gamma}{\sqrt 3}\right)^3. \nonumber 
\label{eq:ov_std_uni}
\end{IEEEeqnarray} 
The SQNR for the standard ADC is 
\begin{IEEEeqnarray} {RCL}
SQNR_{STD}&= & \frac{\sigma_{x}^2}{\sigma_{q,STD}^2+\sigma_{ov,STD}^2}, \label{eq:SQNR_def} \nonumber  \\ \nonumber 
&=& \frac{1}{\frac{\gamma^2}{3\,( 2^{2n})}+\left(1-\frac{\gamma}{\sqrt{3}}\right)^3\,\mathbf{1}_{\gamma \in [0, \sqrt{3}]}}, \nonumber
\label{eq:SQNR_std_uni}
\end{IEEEeqnarray} 
where $\mathbf{1}_{\gamma \in [0, \sqrt{3}]}$ is the indicator function, which reflects the fact that the overload distortion contributes to SQNR only when $\gamma \in [0, \sqrt{3}]$.

\subsection{SQNR for Standard ADC -- Gaussian Distributed Input}
\label{sec:SQNR_der_gau}
Next, consider the input to follow the Gaussian distribution $f(x)=\frac{1}{\sqrt{2\pi \sigma_{x}^2}}\,e^{-\frac{x^2}{2\sigma_{x}^2}},\, x \in \mathbb{R}$, with mean zero and variance $\sigma_{x}^2$. Under the assumption of high-rate quantization, the quantization noise  is approximately uniform and its variance is given by $$\sigma_{q,STD}^2 = \frac{\Delta_{STD}^2}{12}=\frac{V_{ref}^2}{3\,( 2^{2n})}.$$ 
The overload distortion is given by
\begin{IEEEeqnarray} {RCL}
\sigma_{ov,STD}^2 &= &2\int_{V_{ref}}^{\infty}(x-V_{ref})^2\,f_x(x)\,\mathrm{d}x, \nonumber \\
&& \text{(by a change of variable)}  \nonumber \\
&=& \frac{1}{\sqrt{2\pi}}\int_{V_{ref}/\sigma_{x}}^{\infty}(x\sigma_{x}^2-V_{ref})^2\,e^{-x^2/2}\,\mathrm{d}x, \nonumber \\
&=& \underset{I_1}{\underbrace{\frac{\sigma_{x}^2}{\sqrt{2\pi}}\hspace{-1mm}\int_{\gamma}^{\infty}\hspace{-0.3cm}\hspace{-1mm}x^2\,e^{-x^2/2}\,\mathrm{d}x}} - \underset{I_2}{\underbrace{\frac{2V_{ref}\sigma_{x}}{\sqrt{2\pi}}\hspace{-1.5mm}\int_{\gamma}^{\infty}\hspace{-0.4cm}x\,e^{-x^2/2}\,\mathrm{d}x}} \nonumber \\ 
&&+ \underset{I_3}{\underbrace{\frac{V_{ref}^2}{\sqrt{2\pi}}\int_{\gamma}^{\infty}e^{-x^2/2}\,\mathrm{d}x}}.
\label{eq:ov_std_gau2}
\end{IEEEeqnarray}
One can readily verify that
\begin{IEEEeqnarray} {RCL}
I_1&=&\sigma_{x}^2\left(\frac{\gamma}{\sqrt{2\pi}}e^{-\gamma^2/2}+Q(\gamma)\right), \nonumber \\
I_2&=&\sigma_{x}^2\, \gamma\, \sqrt{\frac{2}{\pi}}e^{-\gamma^2/2}, \,\, \text{and} \nonumber \\
I_3&=&V_{ref}^2 Q(\gamma), \nonumber 
\label{eq:integrals_gau}
\end{IEEEeqnarray} 
where $Q(x) = \frac{1}{\sqrt{2\pi}}\int_{x}^\infty e^{-\xi^2/2}\,\mathrm{d}\xi$. Substituting the above expressions for $I_1$, $I_2$, and $I_3$ in \eqref{eq:ov_std_gau2} and simplifying gives
\begin{IEEEeqnarray} {RCL}
\hspace{-0.5 cm}\sigma_{ov,STD}^2 &= &\sigma_{x}^2\underset{\psi(\gamma)}{\left[\underbrace{(1+\gamma^2)Q(\gamma)-\frac{1}{\sqrt{2\pi}}\gamma e^{-\gamma^2/2}}\right]}.
\label{eq:ov_std_gau}
\end{IEEEeqnarray} 
Thus, the SQNR turns out to be
\begin{IEEEeqnarray} {RCL}
SQNR_{STD}&= & \frac{\sigma_{x}^2}{\frac{V_{ref}^2}{3\,( 2^{2n})}+\sigma_x^2 \psi(\gamma)}=\frac{1}{\frac{\gamma^2}{3\,( 2^{2n})}+ \psi(\gamma)}. \nonumber 
\label{eq:SQNR_std_gau}
\end{IEEEeqnarray} 

\subsection{SQNR for Standard ADC -- Laplacian Distributed Input}
\label{sec:SQNR_der_lap}
Consider the input to be Laplacian distributed with mean zero and variance $\sigma_{x}^2=2\lambda^2$. The Laplacian p.d.f. is given by $f(x)=\frac{1}{2\lambda}\,e^{-\frac{|x|}{\lambda}},\, x \in \mathbb{R}$. The quantization noise variance is given by $\sigma_{q,STD}^2 = \frac{\Delta_{STD}^2}{12}=\frac{V_{ref}^2}{3\,( 2^{2n})}$. The overload distortion is given by
\begin{IEEEeqnarray} {RCL}
\sigma_{ov,STD}^2 &= &2\int_{V_{ref}}^{\infty}(x-V_{ref})^2\,f_x(x)\,\mathrm{d}x, \nonumber \\
&=& \frac{1}{\lambda}\int_{V_{ref}}^{\infty}(x-V_{ref})^2\,e^{-|x|/\lambda}\,\mathrm{d}x, \nonumber \\
&=& \underset{J_1}{\underbrace{\frac{1}{\lambda}\int_{V_{ref}}^{\infty}\hspace{-0.3cm}x^2\,e^{-x/\lambda}\,\mathrm{d}x}}\, -  \underset{J_2}{\underbrace{\frac{2V_{ref}}{\lambda}\int_{V_{ref}}^{\infty}\hspace{-0.3cm}x\,e^{-x/\lambda}\,\mathrm{d}x}} \nonumber \\ 
&&+  \underset{J_3}{\underbrace{\frac{V_{ref}^2}{\lambda}\int_{V_{ref}}^{\infty}\hspace{-0.3cm}\,e^{-x/\lambda}\,\mathrm{d}x}}.
\label{eq:ov_std_lap2}
\end{IEEEeqnarray} 
The integrals evaluate to the following:
\begin{IEEEeqnarray} {RCL}
J_1&=&V_{ref}^2e^{-V_{ref}/\lambda}+2\lambda V_{ref}e^{-V_{ref}/\lambda}+2\lambda^2e^{-V_{ref}/\lambda}, \nonumber \\
J_2&=&2V_{ref}^2e^{-V_{ref}/\lambda}+2\lambda V_{ref}e^{-V_{ref}/\lambda}, \,\, \text{and}\nonumber \\
J_3&=&V_{ref}^2e^{-V_{ref}/\lambda}. \nonumber 
\label{eq:integrals_lap}
\end{IEEEeqnarray} 
Substituting for $J_1$, $J_2$, and $J_3$ in \eqref{eq:ov_std_lap2} and simplifying yields a compact expression for the variance of the overload distortion:
\begin{IEEEeqnarray} {RCL}
\sigma_{ov,STD}^2 &= &2\lambda^2 e^{-V_{ref}/\lambda} = \sigma_{x}^2\, e^{-\sqrt{2}\gamma}. \nonumber 
\label{eq:ov_std_lap}
\end{IEEEeqnarray} 
\indent Correspondingly, the SQNR is 
\begin{IEEEeqnarray} {RCL}
SQNR_{STD}&= & \frac{\sigma_{x}^2}{\frac{V_{ref}^2}{3\,( 2^{2n})}+\sigma_{x}^2\, e^{-\sqrt{2}\gamma}}, \nonumber \\
&=&\frac{1}{\frac{\gamma^2}{3\,( 2^{2n})}+ e^{-\sqrt{2}\gamma}}.  \nonumber 
\label{eq:SQNR_std_lap}
\end{IEEEeqnarray} 

\bibliographystyle{IEEEtran}
\bibliography{refs_thesis.bib}

\end{document}